# Taylor expansion method to solve Dokshitzer-Gribov-Lipatov-Altarelli-Parisi equations in leading and next-to-leading orders at small-x

R Rajkhowa[1] and J K Sarma[2]

Physics Department, Tezpur University, Napam, Tezpur-784 028, Assam, India

[1]E-mail: rasna@tezu.ernet.in, [2]E-mail: jks@tezu.ernet.in

## Abstract

We present particular and unique solutions of singlet and non-singlet Dokshitzer-Gribov-Lipatov-Altarelli-Parisi (DGLAP) evolution equations in leading order (LO) and next-to-leading order (NLO) and gluon, sea and valence quark Dokshitzer-Gribov-Lipatov-Altarelli-Parisi (DGLAP) evolution equations in leading order (LO) by applying Taylor expansion method at small-*x*. We obtain *t*-evolutions of deuteron, proton, neutron, difference and ratio of proton and neutron, gluon, light sea and valence quark structure functions and *x*-evolution of deuteron, gluon and light sea quark structure functions at small-*x* from DGLAP evolution equations. The results of *t*-evolutions are compared with HERA and NMC low-*x* low-$Q^2$ data and *x*-evolutions are compared with NMC low-*x* low-$Q^2$ data and recent global parameterization. And also we compare our results of *t*-evolution of proton structure functions with a recent global parameterization.

## 1. Introduction

The Dokshitzer-Gribov-Lipatov-Altarelli-Parisi (DGLAP) evolution equations [1-4] are fundamental tools to study the $t$ ( = ln($Q^2$/Ë$^2$)) and *x*-evolutions of structure functions, where $x$ and $Q^2$ are Bjorken scaling and four momenta transfer in a deep inelastic scattering (DIS) process [5] respectively and L is the QCD cut-off parameter. On the other hand, the study of structure functions at small-*x* has become topical in view [6] of high energy collider and supercollider experiments [7]. Solutions of DGLAP evolution equations give quark and gluon structure functions which produce ultimately proton, neutron and deuteron structure functions. In the literature, there exist various techniques to solve DGLAP evolution equations, such as the so called brute-force method [8], use of Laguerre polynomials [9, 10], and solution in Mellin moment space [11] with subsequent inversion. The shortcomings common to all are the computer time required and decreasing accuracy for $x \to 0$ [12]. More precise approach is the matrix approach [12] to the solution of the DGLAP evolution equations, yet it is also a numerical solution. Thus though numerical solutions are available in the literature [13], the explorations of the possibility of obtaining analytical solutions of DGLAP evolution equations are always interesting. In this connection, some particular solutions computed from general solutions of DGLAP evolution equations at small-*x* in leading order have already been obtained by applying Taylor expansion method [14] and *t*-evolutions [15] and *x*- evolutions [16] of structure functions have been presented.

The present paper reports particular and unique solutions of DGLAP evolution equations computed from complete solutions in leading order and next-to-leading order at small-*x* and calculation of *t* and *x*-evolutions for singlet, non-singlet, gluon, light sea and valence quark structure functions and hence *t*-evolutions of deuteron, proton, neutron, difference and ratio of proton and neutron, gluon, light sea and valence quark structure functions and *x*-evolutions of deuteron, gluon and light sea quark structure functions. Here, the integro-differential DGLAP evolution equations have been converted into first order partial differential equations by applying Taylor expansion in the small-*x* limit. Then those have been solved by standard analytical methods. In calculating structure functions, input data points have been taken from experimental data directly unlike the usual practice of using an input *x*-distribution function introduced by hand. On the other hand, our method is capable of calculating *x*-evolution of structure functions at least at small-*x* limit. The possible explanation is this: traditionally the DGLAP equations provide a means of calculating the manner in which the parton distributions change at fixed *x* as $Q^2$ varies. This change comes about because of the various types of parton branching emission processes and the *x*-distributions are modified as the initial momentum is shared among the various daughter partons. However the exact rate of modifications of *x*-distributions at fixed $Q^2$ can not be obtained from the DGLAP equations since it depends not only on the initial *x* but also on the rate of change of parton distributions with respect to *x*, $d^nF(x)/dx^n$ ($n$ = 1 to $\infty$), up to infinite order. Physically this implies that at high-*x*, the parton has a large momentum fraction at its disposal and as a result it radiates partons including gluons in innumerable ways, some of them involving complicated QCD mechanisms. However for small-*x*, many of the radiation processes will cease to occur due to momentum constraints and the *x*-evolutions get simplified. It is then possible to visualize a situation in which the modification of the *x*-distribution simply depends on its initial value and its first derivative. In this simplified situation, the DGLAP equations give information on the shapes of the *x*-distribution as demonstrated in this paper. The results of *t*-evolutions are compared with HERA and NMC low-*x* low-$Q^2$ data and *x*-evolution are compared with NMC low-*x* low-$Q^2$ data and recent global parameterization. And also we compare our results of *t*-evolution of proton structure functions with a recent global parameterization. In the present paper, section 1 is the introduction. In section 2 necessary theory has been discussed. Section 3 gives results and discussion, and section 4 is conclusion.

## 2. Theory

The DGLAP evolution equations for singlet, non-singlet, gluon, sea and valence quark [17-21] structure functions are in the standard forms

$$\frac{\partial F_2^S(x,t)}{\partial t} - \frac{\alpha'_s(t)}{2\pi}\left[\frac{2}{3}\{3+4\ln(1-x)\}F_2^S(x,t)+\frac{4}{3}\int_x^1\frac{dw}{1-w}\left\{(1+w^2)F_2^S\left(\frac{x}{w},t\right)-2F_2^S(x,t)\right\}\right.$$

$$\left. + n_f\int_x^1\left\{w^2+(1-w)^2\right\}G\left(\frac{x}{w},t\right)dw\right] = 0, \tag{1}$$

$$\frac{\partial F_2^{NS}(x,t)}{\partial t} - \frac{\alpha'_s(t)}{2\pi}\left[\frac{2}{3}\{3+4\ln(1-x)\}F_2^{NS}(x,t)+\frac{4}{3}\int_x^1\frac{dw}{1-w}\left\{(1+w^2)F_2^{NS}\left(\frac{x}{w},t\right)-2F_2^{NS}(x,t)\right\}\right]=0, \tag{2}$$

$$\frac{\partial G(x,t)}{\partial t} - \frac{3\alpha'_s(t)}{\pi}\left\{\left(\frac{11}{12}-\frac{n_f}{18}+\ln(1-x)\right)G(x,t)+\int_x^1 dw\left[\frac{wG(x/w,t)-G(x,t)}{1-w}\right]+\left(w(1-w)+\frac{1-w}{w}\right)G(x/w,t)\right\}$$

$$+\frac{3\alpha'_s(t)}{\pi}\left\{\int_x^1 dw\left[\frac{2}{9}\left(\frac{1+(1-w)^2}{w}\right)F_2^S(x/w,t)\right]\right\}=0, \tag{3}$$

$$*\ \frac{\partial F_2^{S/}\left(x,Q^2\right)}{\partial\ln Q^2}-\frac{2\alpha'_s(t)}{3\pi}\left[\int_x^1\frac{dw}{1-w}\left\{\left(1+w^2\right)F_2^{S/}\left(x/w,Q^2\right)-2F_2^{S/}\left(x,Q^2\right)\right\}+\frac{3}{2}\left(1+\frac{4\ln(1-x)}{3}\right)F_2^{S/}(x,Q^2)\right]$$

$$+\frac{\alpha'_s(t)}{4\pi}\left[w^2+(1+w)^2\right]G(x/w,Q^2)=0, \tag{4}$$

$$*\ \frac{\partial F_2^{v}\left(x,Q^2\right)}{\partial\ln Q^2}-\frac{2\alpha'_s(t)}{3\pi}\left[\int_x^1\frac{dw}{1-w}\left\{\left(1+w^2\right)F_2^{v}\left(x/w,Q^2\right)-2F_2^{v}\left(x,Q^2\right)\right\}+\frac{3}{2}\left(1+\frac{4\ln(1-x)}{3}\right)F_2^{v}(x,Q^2)\right] \tag{5}$$

for LO. And

$$*\ \frac{\partial F_2^S(x,t)}{\partial t}-\frac{\alpha_s(t)}{2\pi}\left[\frac{2}{3}\{3+4\ln(1-x)\}F_2^S(x,t)+\frac{4}{3}\int_x^1\frac{dw}{1-w}\{(1+w^2)F_2^S\left(\frac{x}{w},t\right)-2F_2^S(x,t)\}\right.$$

$$+n_f\int_x^1\{w^2+(1-w)^2\}G\left(\frac{x}{w},t\right)dw\}]-\left(\frac{\alpha_s(t)}{2\pi}\right)^2[(x-1)F_2^S(x,t)\int_0^1 f(w)dw+\int_x^1 f(w)F_2^S\left(\frac{x}{w},t\right)dw$$

$$+\int_x^1 F_{qq}^S(w)F_2^S\left(\frac{x}{w},t\right)dw+\int_x^1 F_{qg}^S(w)G\left(\frac{x}{w},t\right)dw\,]=0, \tag{6}$$

$$*\ \frac{\partial F_2^{NS}(x,t)}{\partial t}-\frac{\alpha_s(t)}{2\pi}\left[\frac{2}{3}\{3+4\ln(1-x)\}F_2^{NS}(x,t)+\frac{4}{3}\int_x^1\frac{dw}{1-w}\{(1+w^2)F_2^{NS}\left(\frac{x}{w},t\right)-2F_2^{NS}(x,t)\}\right]$$

$$-\left(\frac{\alpha_s(t)}{2\pi}\right)^2\left[(x-1)F_2^{NS}(x,t)\int_0^1 f(w)dw+\int_x^1 f(w)F_2^{NS}\left(\frac{x}{w},t\right)dw\right]=0 \tag{7}$$

for NLO, where $F_2^S$ and $F_2^{NS}$ are some combinations of quarks and antiquarks, $F_2^{S/}(x,Q^2)$ = $xu_s$ or $xd_s$ or $xs_s$, $F_2^v(x,Q^2)$ = $xu_v$ or $xd_v$, $t=\ln\frac{Q^2}{\Lambda^2}$, $\alpha'_s(t)=\frac{4\pi}{\beta_0 t}$, $\alpha_s(t)=\frac{4\pi}{\beta_0 t}\left[1-\frac{\beta_1\ln t}{\beta_0^2 t}\right]$, $\Lambda$ is the QCD scale parameter which depends on the renormalization scheme, $\beta_0$ and $\beta_1$ are the expansion coefficient of the $\beta$-function and they are given by $\beta_0=\frac{33-2n_f}{3}$, $\beta_1=\frac{306-38n_f}{3}$, $n_f$ being the number of flavours. Here,

$$f(w)=C_F^2\left[P_F(w)-P_A(w)\right]+\frac{1}{2}C_F C_A\left[P_G(w)+P_A(w)\right]+C_F T_R n_f P_{N_f}(w),$$

$$F_{qq}^S(w)=2C_F T_R n_f F_{qq}(w),\quad F_{qg}^S(w)=C_F T_R n_f F_{qg}^1(w)+C_G T_R n_f F_{qg}^2(w).$$

The explicit forms of higher order kernels are [19-20]

$$P_F(w)=-\frac{2\left(1+w^2\right)}{1-w}\ln w\ln(1-w)-\left(\frac{3}{1-w}+2w\right)\ln w-\frac{1}{2}(1+w)\ln^2 w-5(1-w),$$

$$P_G(w)=\frac{1+w^2}{1-w}\left(\ln^2 w+\frac{11}{3}\ln w+\frac{67}{9}-\frac{\pi^2}{3}\right)+2(1+w)\ln w+\frac{40}{3}(1-w),$$

$$P_{N_F}(w)=\frac{2}{3}\left[\frac{1+w^2}{1-w}\left(-\ln w-\frac{5}{3}\right)-2(1-w)\right],$$

$$P_A(w)=\frac{2(1+w^2)}{1+w}\int_{w/(1+w)}^{1/(1+w)}\frac{dk}{k}\ln\frac{1-k}{k}+2(1+w)\ln w+4(1-w),$$

$$F_{qq}(w)=\frac{20}{9w}-2+6w-\frac{56}{9}w^2+\left(1+5w+\frac{8}{3}w^2\right)\ln w-(1+w)\ln^2 w,$$

$$F_{qg}^1(w)=4-9w-(1-4w)\ln w-(1-2w)\ln^2 w+4\ln(1-w)+\left[2\ln^2\left(\frac{1-w}{w}\right)-4\ln\left(\frac{1-w}{w}\right)-\frac{2}{3}\pi^2+10\right]P_{qg}(w)$$ and

$$F_{qg}^2(w)=\frac{182}{9}+\frac{14}{9}w+\frac{40}{9w}+\left(\frac{136}{3}w-\frac{38}{3}\right)\ln w-4\ln(1-w)-(2+8w)\ln^2 w+$$

$$\left[-\ln^2 w+\frac{44}{3}\ln w-2\ln^2(1-w)+4\ln(1-w)+\frac{\pi^2}{3}-\frac{218}{3}\right]P_{qg}(w)+2P_{qg}(-w)\int_{w/1+w}^{1/1+w}\frac{dz}{z}\ln\frac{1-z}{z},$$

where

$P_{qg}(w)=w^2+(1-w)^2$, $C_A$, $C_G$, $C_F$, and $T_R$ are constants associated with the color SU(3) group and $C_A=C_G=N_C=3$, $C_F=(N_C^2-1)/2N_C$ and $T_R=1/2$. $N_C$ is the number of colours.

Let us introduce the variable $u=1-w$ and note that [22]

$$\frac{x}{w}=\frac{x}{1-u}=x+\frac{xu}{1-u}. \tag{8}$$

The series (8) is convergent for $|u|<1$. Since $x<w<1$, so $0<u<1-x$ and hence the convergence criterion is satisfied. Now, using Taylor expansion method [14, 23] we can rewrite $F_2^S(x/w, t)$ as

$$F_2^S(x/w,t)=F_2^S\left((x+\frac{xu}{1-u}),t\right)$$

$$=F_2^S(x,t)+x.\frac{u}{1-u}\frac{\partial F_2^S(x,t)}{\partial x}+\frac{1}{2}x^2\left(\frac{u}{1-u}\right)^2\frac{\partial^2 F_2^S(x,t)}{\partial x^2}+.....$$

which covers the whole range of $u$, $0<u<1-x$. Since $x$ is small in our region of discussion, the terms containing $x^2$ and higher powers of $x$ can be neglected as our first approximation as discussed in our earlier works [15-16, 24-28]. $F_2^S(x/w, t)$ can then be approximated for small-$x$ as

$$F_2^S(x/w,t)\cong F_2^S(x,t)+x.\frac{u}{1-u}\frac{\partial F_2^S(x,t)}{\partial x}. \tag{9}$$

Similarly, $G(x/w, t)$ can be approximated for small-$x$ as

$$G(x/w,t)\cong G(x,t)+x.\frac{u}{1-u}\frac{\partial G(x,t)}{\partial x}. \tag{10}$$

Using equations (9) and (10) in equation (1) and performing $u$-integrations we get

$$\frac{\partial F_2^S(x,t)}{\partial t} - \frac{\alpha_S'(t)}{2\pi}\left[A_1(x)F_2^S(x,t) + A_2(x)G(x,t) + A_3(x)\frac{\partial F_2^S(x,t)}{\partial x} + A_4(x)\frac{\partial G(x,t)}{\partial x}\right] = 0. \tag{11}$$

Here

$$A_1(x) = \frac{2}{3}\{3 + 4\ln(1-x) + (x-1)(x+3)\},\quad A_2(x) = n_f\left[\frac{1}{3}(1-x)(2-x+2x^2)\right],$$

$$A_3(x) = \frac{2}{3}\{x(1-x^2) + 2x\ln(\frac{1}{x})\},\qquad A_4(x) = n_f\, x\{\ln\frac{1}{x} - \frac{1}{3}(1-x)(5-4x+2x^2)\}.$$

We assume [13, 15-16, 24-28]

$$G(x, t) = K(x)\, F_2^S(x, t). \tag{12}$$

Here, $K$ is a function of $x$. It is to be noted that if we consider Regge behaviour of singlet and gluon structure function, it is possible to solve coupled evolution equations for singlet and gluon structure functions and evaluate $K(x)$ [29] in LO and NLO. Otherwise this is a parameter to be estimated from experimental data. We take $K(x) = k,\ ax^b,\ ce^{-dx}$, where $k$, $a$, $b$, $c$, $d$ are constants. Therefore equations (11) becomes

$$\frac{\partial F_2^S(x,t)}{\partial t} - \frac{A_f}{t}\left[L_1(x)F_2^S(x,t) + L_2(x)\frac{\partial F_2^S(x,t)}{\partial x}\right] = 0. \tag{13}$$

Here,

$$L_1(x) = A_1(x) + K(x)A_2(x) + A_4(x)\frac{\partial K(x)}{\partial x},\quad L_2(x) = A_3(x) + K(x)\, A_4(x) \text{ and } A_f = 4/(33-2n_f).$$

The general solution [23, 30] of equation (13) is $F(U, V) = 0$, where $F$ is an arbitrary function and $U(x, t, F_2^S) = C_1$ and $V(x, t, F_2^S) = C_2$, where $C_1$ and $C_2$ are constants and they form a solutions of equations

$$\frac{dx}{A_f L_2(x)} = \frac{dt}{-t} = \frac{dF_2^S(x,t)}{-A_f L_1(x) F_2^S(x,t)}. \tag{14}$$

Solving equation (14) we obtain

$$U\left(x,t,F_2^S\right) = t\exp\left[\frac{1}{A_f}\int\frac{1}{L_2(x)}dx\right] \text{ and } V\left(x,t,F_2^S\right) = F_2^S(x,t)\exp\left[\int\frac{L_1(x)}{L_2(x)}dx\right].$$

## 2. (a) Complete and Particular Solutions

Since $U$ and $V$ are two independent solutions of equation (14) and if $\alpha$ and $\beta$ are arbitrary constants, then $V = \alpha U + \beta$ may be taken as a complete solution [23, 30] of equation (13). We take this form as this is the simplest form of a complete solution which contains both the arbitrary constants $\alpha$ and $\beta$. Earlier [13, 15-16, 24] we considered a solution AU + BV = 0, where A and B are arbitrary constants. But that is not a complete solution having both the arbitrary constants as this equation can be transformed to the form V = CU, where C=-A/B, i. e, the equation contains only one arbitrary constant. So, the complete solution

$$F_2^S(x,t)\exp\left[\int \frac{L_1(x)}{L_2(x)}dx\right] = \alpha t \exp\left[\frac{1}{A_f}\int \frac{L_1(x)}{L_2(x)}dx\right] + \beta \tag{15}$$

is a two-parameter family of surfaces, which does not have an envelope, since the arbitrary constants enter linearly [23]. Differentiating equation (15) with respect to $\beta$ we get 0 = 1, which is absurd. Hence there is no singular solution. The one parameter family determined by taking $\beta = \alpha^2$ has equation

$$F_2^S(x,t)\exp\left[\int \frac{L_1(x)}{l_2(x)}dx\right] = \alpha t \exp\left[\frac{1}{A_f}\int \frac{1}{L_2(x)}dx\right] + \alpha^2. \tag{16}$$

Differentiating equation (16) with respect to $\alpha$, we get $\alpha = -\frac{1}{2}t\exp\left[\frac{1}{A_f}\int \frac{1}{L_2(x)}dx\right]$. Putting the value of $\alpha$ in equation (16), we get

$$F_2^S(x,t) = -\frac{1}{4}t^2\exp\left[\int\left(\frac{2}{A_f L_2(x)} - \frac{L_1(x)}{L_2(x)}\right)dx\right], \tag{17}$$

which is merely a particular solution of the general solution. Now, defining

$F_2^S(x,t_0) = -\frac{1}{4}t_0^2\exp\left[\int\left(\frac{2}{A_f L_2(x)} - \frac{L_1(x)}{L_2(x)}\right)dx\right]$, at $t = t_0$, where, $t_0 = \ln(Q_0^2/\Lambda^2)$ at any lower value $Q = Q_0$, we get from equation (17)

$$F_2^S(x,t) = F_2^S(x,t_0)\left(\frac{t}{t_0}\right)^2, \tag{18}$$

which gives the $t$-evolution of singlet structure function $F_2^S(x, t)$. Again defining,

$F_2^S(x_0,t) = -\frac{1}{4}t^2\exp\left[\int\left(\frac{2}{A_f L_2(x)} - \frac{L_1(x)}{L_2(x)}\right)dx\right]_{x = x_0}$, we obtain from equation (17)

$$F_2^S(x,t) = F_2^S(x_0,t)\exp\left[\int_{x_0}^{x}\left(\frac{2}{A_f L_2(x)} - \frac{L_1(x)}{L_2(x)}\right)dx\right] \tag{19}$$

which gives the $x$-evolution of singlet structure function $F_2^S(x, t)$. Proceeding in the same way, we get

$$F_2^R = F_2^R(x,t_0)\left(\frac{t}{t_0}\right)^2, \text{ and } G(x,t) = G(x,t_0)\left(\frac{t}{t_0}\right)^2 \tag{20}$$

which give the $t$-evolutions of non-singlet, light sea and valence quark, and gluon structure functions in LO. Here $R$ represents $NS$, $S'$, $v$ and also

$$F_2^R(x,t) = F_2^R(x_0,t)\exp\left[\int_{x_0}^{x}\left(\frac{2}{A_f Q(x)} - \frac{P(x)}{Q(x)}\right)\right]dx, \tag{21}$$

$$G(x,t) = G(x_0,t)\exp\left[\int_{x_0}^{x}\left(\frac{2}{A'_f Q_1(x)} - \frac{P_1(x)}{Q_1(x)}\right)dx\right], \tag{22}$$

which give the $x$-evolutions of non-singlet, light sea and valence quark, and gluon structure functions in LO.

Proceeding exactly in the same way, from equations (6) and (7) we get

$$F_2^{S,NS}(x,t) = F_2^{S,NS}(x,t_0)\left(\frac{t^{(b/t+1)}}{t_0^{(b/t_0+1)}}\right)^2 exp\left[2b\left(\frac{1}{t}-\frac{1}{t_0}\right)\right], \tag{23}$$

$$F_2^S(x,t)=F_2^S(x_0,t)\exp\int_{x_0}^{x}\left[\frac{2}{a}.\frac{1}{L_2(x)+T_0M_2(x)}-\frac{L_1(x)+T_0M_1(x)}{L_2(x)+T_0M_2(x)}\right]dx, \tag{24}$$

$$F_2^{NS}(x,t)=F_2^{NS}(x_0,t)\exp\int_{x_0}^{x}\left[\frac{2}{a}.\frac{1}{A_3(x)+T_0B_3(x)}-\frac{A_1(x)+T_0B_1(x)}{A_3(x)+T_0B_3(x)}\right]dx, \tag{25}$$

which gives the $t$ and $x$-evolution of singlet and non-singlet structure functions in NLO where $a = 2/\beta_0$, $b = \beta_1/\beta_0^2$. Here for possible solutions in NLO, we are to put an extra assumption [21, 27-28] $\left(\frac{\alpha_s(t)}{2\pi}\right)^2 = T_0\left(\frac{\alpha_s(t)}{2\pi}\right)$, where $T_0$ is a numerical parameter. By a suitable choice of $T_0$ we can reduce the error to a minimum. We observe that in case of $t$-evolutions, if $b$ tends to zero, then equation (23) tends to equation (18) and (20) respectively, i.e., solution of NLO equations goes to that of LO equations. Physically $b$ tends to zero means number of flavours is high. Here

$$M_1(x) = x\int_0^1 f(w)dw - \int_0^x f(w)dw + \frac{4}{3}N_f\int_x^1 F_{qq}(w)dw + K(x)\int_x^1 F_{qg}^S(w)dw + x\int_x^1\frac{1-w}{w}F_{qg}^S(w)dw\frac{\partial K(x)}{\partial x},$$

$$M_2(x) = x\int_x^1\{f(w)+\frac{4}{3}N_f F_{qq}(w)\}\frac{1-w}{w}dw + K(x)x\int_x^1\frac{1-w}{w}F_{qg}^S(w)dw,$$

$$B_1(x) = -\int_0^x f(w)dw + x\int_0^1 f(w)dw, \qquad B_2(x) = [\frac{1}{4}(1-x)(2-x+2x^2)].$$

$$B_3(x) = x\int_x^1\frac{1-w}{w}f(w)dw, \qquad B_4(x) = x\left[-\frac{1}{4}(1-x)(5-4x+2x^2)+\frac{3}{4}\ln(\frac{1}{x}\right].$$

For non-singlet and light valence quark structure functions

$$P(x) = \frac{3}{2}A_1(x),\ \ Q(x) = \frac{3}{2}A_3(x),$$

and for light sea quark structure functions

$$P(x) = \frac{3}{2}A_1(x) + K(x)B_2(x) + B_4(x)\frac{\partial K(x)}{\partial x},\ \ Q(x) = \frac{3}{2}A_3(x) + K(x)B_2(x),$$

$$P_1(x) = -\left[\frac{2}{9}(1-x)+\frac{1}{9}(1-x)^2+\frac{5}{9}\ln x\right]K_1(x) + x\left[\frac{4}{9x}+\frac{4}{9}(1-x)+\frac{1}{9}(1-x)^2+\frac{8}{9}\ln x-\frac{4}{9}\right]\frac{\partial K_1(x)}{\partial x}$$

$$+\left(\frac{11}{12}-\frac{n_f}{18}\right)+\ln(1-x)-\left[2(1-x)-\frac{1}{2}(1-x)^2+\frac{1}{3}(1-x)^3+\ln x\right],$$

$$Q_1(x) = x\left[\frac{4}{9x}+\frac{4}{9}(1-x)+\frac{1}{9}(1-x)^2+\frac{8}{9}\ln x-\frac{4}{9}\right]K_1(x)+x\left[\frac{1}{x}+2(1-x)+\frac{1}{3}(1-x)^3+2\ln x-1\right],$$

$$A'_f = 9A_f,\ K_1(x) = 1/K(x).$$

For all these particular solutions, we take $b = a^2$. But if we take $b = a$ and differentiate with respect to $a$ as before, we can not determine the value of $a$. In general, if we take $b = a^y$, we get in the solutions, the powers of $(t/t_0)$ and the numerators of the first term inside the integral sign be $y/(y-1)$ for $t$ and $x$-evolutions respectively in LO. Similarly we get the powers of $t^{b/t+1}\big/t_0^{\,b/t_0+1}$ and co-efficient of $b(1/t-1/t_0)$ of exponential part in $t$-evolutions and the numerators of the first term inside the integral sign be $y/(y-1)$ for $x$-evolutions in NLO. Then if $y$ varies from minimum (=2) to maximum (=∞) then $y/(y-1)$ varies from 2 to 1.

For phenomenological analysis, we compare our results with various experimental structure functions. Deuteron, proton and neutron structure functions [5] can be written as

$F_2^d(x, t) = (5/9)\ F_2^S(x, t)$, (26)

$F_2^p(x, t) = (5/18)\ F_2^S(x, t) + (3/18)\ F_2^{NS}(x, t)$, (27)

$F_2^n(x, t) = (5/18)\ F_2^S(x, t) - (3/18)\ F_2^{NS}(x, t)$. (28)

Now using equations (18), (19), (20) and (23), (24) in equations (26), (27) and (28) we will get $t$-evolutions of deuteron, proton, neutron, difference and ratio of proton and neutron and $x$-evolution of deuteron structure functions at small-$x$ as

$$F_2^{d,p,n}(x,t) = F_2^{d,p,n}(x,t_0)\left(\frac{t}{t_0}\right)^2, \tag{29}$$

$$F_2^p(x,t)-F_2^n(x,t) = [F_2^p(x,t_0)-F_2^n(x,t_0)](\frac{t}{t_0})^2, \tag{30}$$

$$\frac{F_2^p(x,t)}{F_2^n(x,t)} = \frac{F_2^p(x,t_0)}{F_2^n(x,t_0)} = R(x), \tag{31}$$

$$F_2^d(x,t) = F_2^d(x_0,t)\exp\left[\int_{x_0}^{x}\left(\frac{2}{A_f L_2(x)}-\frac{L_1(x)}{L_2(x)}\right)dx\right] \tag{32}$$

in LO for $b = a^2$. The corresponding results in NLO for $b = a^2$ are

$$F_2^{d,p,n}(x,t) = F_2^{d,p,n}(x,t_0)\left(\frac{t^{(b/t+1)}}{t_0^{(b/t_0+1)}}\right)^2 \exp\left[2b\left(\frac{1}{t}-\frac{1}{t_0}\right)\right], \tag{33}$$

$$F_2^p(x,t)-F_2^n(x,t) = [F_2^p(x,t_0)-F_2^p(x,t_0)]\left(\frac{t^{(b/t+1)}}{t_0^{(b/t_0+1)}}\right)^2 \exp\left[2b\left(\frac{1}{t}-\frac{1}{t_0}\right)\right], \tag{34}$$

$$\frac{F_2^p(x,t)}{F_2^n(x,t)} = \frac{F_2^p(x,t_0)}{F_2^n(x,t_0)} = R(x), \tag{35}$$

$$F_2^d(x,t)=F_2^d(x_0,t)\exp\int_{x_0}^{x}\left[\frac{2}{a}\cdot\frac{1}{L_2(x)+T_0M_2(x)}-\frac{L_1(x)+T_0M_1(x)}{L_2(x)+T_0M_2(x)}\right]dx, \tag{36}$$

where $R(x)$ is a constant for fixed-$x$. The determination of $x$-evolutions of proton and neutron structure functions like that of deuteron structure function is not suitable by this methodology; because to extract the $x$-evolution of proton and neutron structure functions, we are to put equations (19) and (21) in equations (27) and (28). But as the functions inside the integral sign of equations (19) and (21) are different, we need to separate the input functions $F_2^S(x_0, t)$ and $F_2^{NS}(x_0, t)$ from the data points to extract the $x$-evolutions of the proton and neutron structure functions, which may contain large errors.

## 2. (b) Unique Solutions

Due to conservation of the electromagnetic current, $F_2$ must vanish as $Q^2$ goes to zero [5, 31]. Also $R\rightarrow 0$ in this limit. Here $R$ indicates ratio of longitudinal and transverse cross-sections of virtual photon in DIS process. This implies that scaling should not be a valid concept in the region of very low-$Q^2$. The exchanged photon is then almost real and the close similarity of real photonic and hadronic interactions justifies the use of the Vector Meson Dominance (VMD) concept [32-33] for the description of $F_2$. In the language of perturbation theory, this concept is equivalent to a statement that a physical photon spends part of its time as a 'bare', point-like photon and part as a virtual hadron [31]. The power and beauty of explaining scaling violations with field theoretic methods (i.e., radiative corrections in QCD) remains, however, unchallenged in as much as they provide us with a framework for the whole $x$-region with essentially only one free parameter $\Lambda$ [34]. For $Q^2$ values much larger than $\Lambda^2$, the effective coupling is small and a perturbative description in terms of quarks and gluons interacting weakly makes sense. For $Q^2$ of order $\Lambda^2$, the effective coupling is infinite and we cannot make such a picture, since quarks and gluons will arrange themselves into strongly bound clusters, namely, hadrons [5] and so the perturbation series breaks down at small-$Q^2$ [5]. Thus, it can be thought of $\Lambda$ as marking the boundary between a world of quasi-free quarks and gluons, and the world of pions, protons, and so on. The value of $\Lambda$ is not predicted by the theory; it is a free parameter to be determined from experiment. It should expect that it is of the order of a typical hadronic mass [5]. Since the value of $\Lambda$ is so small we can take at $Q = \Lambda$, $F_2^S(x, t) = 0$ due to conservation of the electromagnetic current [31]. This dynamical prediction agrees with most ad hoc parameterizations and with the data [34]. Using this boundary condition in equation (15) we get $\beta = 0$ and

$$F_2^S(x,t)=\alpha t\exp\left[\int\left(\frac{1}{A_f L_2(x)}-\frac{L_1(x)}{L_2(x)}\right)dx\right]. \tag{37}$$

Now, defining $F_2^S(x,t_0)=\alpha t_0\exp\left[\int\left(\frac{1}{A_f L_2(x)}-\frac{L_1(x)}{L_2(x)}\right)dx\right]$, at $t = t_0$, where $t_0 = \ln(Q_0^2/\Lambda^2)$ at any lower value $Q = Q_0$, we get from equations (37)

$$F_2^S(x,t)=F_2^S(x,t_0)\left(\frac{t}{t_0}\right), \tag{38}$$

which gives the $t$-evolutions of singlet structure function $F_2^S(x, t)$ in LO. Proceeding in the same way, we get

$$F_2^S(x,t)=F_2^S(x_0,t)exp\left[\int_{x_0}^{x}\left(\frac{1}{A_f L_2(x)}-\frac{L_1(x)}{L_2(x)}\right)dx\right] \tag{39}$$

which gives the $x$-evolutions of singlet structure function $F_2^S(x, t)$ in LO. Similarly, we get for non-singlet, light sea and valence quark, and gluon structure functions

$$F_2^R = F_2^R(x,t_0)\left(\frac{t}{t_0}\right), \quad G(x,t)=G_0(x,t_0)\left(\frac{t}{t_0}\right), \tag{40}$$

$$F_2^R(x,t)=F_2^R(x_0,t)\exp\left[\int_{x_0}^{x}\left(\frac{1}{A_f Q(x)}-\frac{P(x)}{Q(x)}\right)\right]dx, \tag{41}$$

$$G(x,t)=G(x_0,t)\exp\left[\int_{x_0}^{x}\left(\frac{1}{A'_f Q_1(x)}-\frac{P_1(x)}{Q_1(x)}\right)dx\right], \tag{42}$$

which give the $t$ and $x$-evolutions of non-singlet, light sea and valence quark, and gluon structure functions in LO and

$$F_2^{S,NS}(x,t)=F_2^{S,NS}(x,t_0)\left(\frac{t^{(b/t+1)}}{t_0^{(b/t_0+1)}}\right)\exp\left[b\left(\frac{1}{t}-\frac{1}{t_0}\right)\right], \tag{43}$$

$$F_2^S(x,t)=F_2^S(x_0,t)exp\int_{x_0}^{x}\left[\frac{1}{a}.\frac{1}{L_2(x)+T_0M_2(x)}-\frac{L_1(x)+T_0M_1(x)}{L_2(x)+T_0M_2(x)}\right]dx, \tag{44}$$

$$F_2^{NS}(x,t)=F_2^{NS}(x_0,t)\exp\int_{x_0}^{x}\left[\frac{1}{a}.\frac{1}{A_5(x)+T_0B_5(x)}-\frac{A_6(x)+T_0B_6(x)}{A_5(x)+T_0B_5(x)}\right]dx, \tag{45}$$

which give the $t$ and $x$-evolutions of singlet and non-singlet structure functions in NLO.

Therefore corresponding results for $t$-evolution of deuteron, proton, neutron, difference and ratio of proton and neutron structure functions are

$$F_2^{d,p,n}(x,t)=F_2^{d,p,n}(x,t_0)\left(\frac{t}{t_0}\right), \tag{46}$$

$$F_2^p(x,t)-F_2^n(x,t)=[F_2^p(x,t_0)-F_2^p(x,t_0)]\left(\frac{t}{t_0}\right), \tag{47}$$

$$\frac{F_2^p(x,t)}{F_2^n(x,t)}=\frac{F_2^p(x,t_0)}{F_2^n(x,t_0)}=R(x) \tag{48}$$

in LO and

$$F_2^{d,p,n}(x,t)=F_2^{d,p,n}(x,t_0)\left(\frac{t^{(b/t+1)}}{t_0^{(b/t_0+1)}}\right)\exp\left[b\left(\frac{1}{t}-\frac{1}{t_0}\right)\right], \tag{49}$$

$$F_2^p(x,t) - F_2^n(x,t) = [F_2^p(x,t_0) - F_2^p(x,t_0)]\left(\frac{t^{(b/t+1)}}{t_0^{(b/t_0+1)}}\right)\exp\left[b\left(\frac{1}{t}-\frac{1}{t_0}\right)\right], \tag{50}$$

$$\frac{F_2^p(x,t)}{F_2^n(x,t)} = \frac{F_2^p(x,t_0)}{F_2^n(x,t_0)} = R(x) \tag{51}$$

in NLO. Again $x$-evolution of deuteron structure function in LO and NLO respectively are

$$F_2^d(x,t) = F_2^d(x_0,t)exp\left[\int_{x_0}^{x}\left(\frac{1}{A_f L_2(x)} - \frac{L_1(x)}{L_2(x)}\right)dx\right], \tag{52}$$

$$F_2^d(x,t) = F_2^d(x_0,t)exp\int_{x_0}^{x}\left[\frac{1}{a}.\frac{1}{L_2(x)+T_0M_2(x)} - \frac{L_1(x)+T_0M_1(x)}{L_2(x)+T_0M_2(x)}\right]dx. \tag{53}$$

Already we have mentioned that the determination of $x$-evolutions of proton and neutron structure functions like that of deuteron structure function is not suitable by this methodology. It is to be noted that unique solutions of evolution equations of different structure functions are same with particular solutions for $y$ maximum ($y = \infty$) in $b = a^y$ relation. The procedure we follow is to begin with input distributions inferred from experiment and to integrate the evolution equations (22), (23), (34) and (38) numerically.

## 3. Results and Discussion

In the present paper, we compare our results of $t$-evolution of deuteron, proton, neutron and difference and ratio of proton and neutron structure functions with the NMC [35] and HERA [36] low-$x$ and low-$Q^2$ data and gluon, sea and valence quark structure functions with recent global parameterizations [37-38], and results of $x$-evolution of deuteron, gluon and sea quark structure functions with NMC low-$x$ and low-$Q^2$ data and recent global parameterization [35, 37]. In case of HERA data, proton and neutron structure functions are measured in the range $2 \le Q^2 \le 50$ GeV$^2$. Moreover, here $P_T \le 200$ MeV, where $P_T$ is the transverse momentum of the final state baryon. In case of NMC data, proton and neutron structure functions are measured in the range $0.75 \le Q^2 \le 27$ GeV$^2$. We consider number of flavours $n_f = 4$. We also compare our results of $t$-evolution of proton structure functions with recent global parameterization [37]. These parameterizations include data from H1-96 / 99, ZEUS-96/97(X0.98), NMC, E665 data.

In fig.1(a-d), we present our results of t-evolutions of deuteron, proton, neutron and difference of proton and neutron structure functions (solid lines) for the representative values of $x$ given in the figures for $y = 2$ (upper solid lines) and $y$ maximum (lower solid lines) in $b = a^y$ relation in NLO. Data points at lowest-$Q^2$ values in the figures are taken as input to test the evolution equation. Agreement with the data [35-36] is good. In the same figures, we also plot the results of $t$-evolutions of deuteron, proton, neutron and difference of proton and neutron structure functions (dashed lines) for the particular solutions in LO. Here, upper dashed lines for $y = 2$ and lower dashed lines for $y$ maximum in $b = a^y$ relation. We observe that $t$-evolutions are slightly steeper in LO calculations than those of NLO. But differences in results for proton and neutron structure functions are smaller and NLO results for $y = 2$ are of better agreement with experimental data in general.

In fig.2, we compare our results of $t$-evolutions of proton structure functions $F_2^p$ (solid lines) with recent global parameterization [37] (long dashed lines) for the representative values of $x$ given in the figures for $y = 2$ (upper solid lines) and $y$ maximum (lower solid lines) in $b = a^y$ relation in NLO. Data points at lowest-$Q^2$ values in the figures are taken as input to test the evolution equation. In the same figure, we also plot the results of $t$-evolutions of proton structure functions $F_2^p$ (dashed lines) for the particular solutions in LO. Here, upper dashed lines for $y = 2$ and lower dashed lines for $y$ maximum in $b = a^y$ relation. We observe that $t$-evolutions are slightly steeper in LO calculations than those of NLO. Agreement with the NLO results is found to be better than with the LO results.

In fig.3(a-e), we present our results of $t$-evolutions of gluon, light sea and valence quark structure functions qualitatively for the representative values of $x$ given in the figures for $y$ minimum ($y = 2$) (upper solid lines and upper dashed lines for gluon and solid lines for light sea and valence sea quark structure functions) and $y$ maximum (lower solid lines and lower dashed lines for gluon and dashed lines for light sea and valence sea quark structure functions) in $b = a^y$ relation. We have taken arbitrary inputs from recent global parameterizations MRST2001 (solid lines in fig 3(a)) and MRST2001J (dashed lines in fig.3 (a)) [37] and MRS data (in fig.3 (b)) [38] at $Q_0^2 = 1$ GeV$^2$ [37] and $Q_0^2 = 4$ GeV$^2$ [38] respectively, and MRST2001 [37] at $Q_0^2 = 1$ GeV$^2$ (in figs. 3(c-e)). It is clear from figures that t-evolutions of gluon, light sea and valence quark structure functions depend upon input $G(x, t_0)$, $F_2^{S'}(x, t_0)$ and $F_2^v(x, t_0)$ values.

Unique solutions of $t$-evolution for structure functions are same with particular solutions for $y$ maximum ($y =$ ) in $b = a^y$ relation in LO and NLO.

In figs.4(a-b), we present our results of $x$-distribution of deuteron structure functions $F_2^d$ in LO (fig.4(a)) for $K(x) = k$ (constant) (solid lines), $K(x) = ax^b$ (dashed lines) and for $K(x) = ce^{-dx}$ (dotted lines), and in NLO (fig.4(b)) for $K(x) = ax^b$ (solid lines) and for $K(x) = ce^{-dx}$ (dashed lines) where $a$, $b$, $c$ and $d$ are constants and for representative values of $Q^2$ given in each figure , and compare them with NMC deuteron low-$x$ low-$Q^2$ data [35]. In each data point for $x$-value just below 0·1 has been taken as input $F_2^d(x_0, t)$. In case of LO, agreement of the results with experimental data is good at $k = 4.5$, $a = 4.5$, $b = 0.01$, $c = 5$, $d = 1$. For $x$-evolutions of deuteron structure function, results of unique solutions and results of particular solutions have not any significance difference in LO [26]. In case of NLO, agreement of the result with experimental data is found to be excellent at $a =10$, $b = 0.016$, $c = 0.5$, $d =-3.8$ for $y$ minimum ($y = 2$) and $a =5.5$, $b = 0.016$, $c = 0.28$, $d =-3.8$ for $y$ maximum ($y = \infty$) in relation $b = a^y$. But agreement of the results with experimental data is found to be very poor for any constant value of $k$. Therefore we do not present our result at $K(x) = k$ in NLO. In fig.5(a-e), we present the sensitivity of our results for $k, a, b, c, d$ in the relation $b = a^y$, for $y$ minimum in LO. If the values of $k$, $a$, $c$, or $d$, respectively are increased, the curves shift upward and if the values of $k$, $a$, $c$, or $d$, respectively are decreased, the curves move in the opposite direction. On the other hand if values of $b$ increased or decreased the curve goes downward or upward directions respectively. In fig.6(a-d), we present the sensitivity of our results for $a, b, c, d$ in the relation $b = a^y$, for $y$ minimum in NLO. If the absolute values of $a$, $b$, $c$, or $d$, respectively are increased, the curves shift upward and if the absolute values of $a$, $b$, $c$, or $d$, respectively are decreased, the curves move in the opposite direction.

In fig.7(a-b), we present the sensitivity of our results in deuteron structure function for different values of $T_0$ at best fit of $K(x) = ax^b$ and $K(x) = ce^{-dx}$ in the relation $b = a^y$, for $y$ minimum and for representative values of $Q^2$ given in each figure. Here, $a = 10$, $b = 0.016$, $c = 0.5$, $d = -3.8$. We observed that if the value of $T_0$ is increased, the curve moves slightly upward and if the value of $T_0$ is decreased, the curve moves slightly downward direction. But the nature of the curve remains same and difference between the curves is extremely small in both cases in the $T_0$ range mentioned in the figure.

In fig.8, we plot $T(t)^2$ and $T_0T(t)$, where $T(t)$ = $á_s$(t)/2ð against $Q^2$ in the $Q^2$ range $0 \leq Q^2 \leq 30$ GeV$^2$ as required by our data used. Though the explicit value of $T_0$ is not necessary in calculating $t$-evolution of, yet we observe that for $T_0 = 0.027$, errors become minimum in the $Q^2$ range $0 \leq Q^2 \leq 30$ GeV$^2$.

In fig.9(a-b), we present our results of $x$-distribution of gluon structure functions for $K_1(x) = ax^b$ , where $a$ and $b$ are constants for representative values of $Q^2$ given in each figure, and compare them with recent global parameterization [37] in the relation $b = a^y$ for $y$ minimum (thick solid lines). In fig.9(a), we observed that agreement of the results with parameterization is found to be very poor for any values of $a$ and $b$ at small-$x$ and agreement is found to be good at high-$x$ at $a = 372$ and $b = 4$. In fig.9(b) agreement of the results with parameterizations is found to be good at $a = 135$ and $b = 1.8$. In the same figures we present the sensitivity of our results for different values of $a$ at fixed value $b$ (dashed lines). We observe that if value of $a$ is increased, the curve goes upward direction and if value of $a$ is decreased, the curve goes downward direction. But the nature of the curve is similar. In fig.10(a-b), we present the sensitivity of our results of gluon structure functions for different values of $b$ at fixed value of $a$ (dashed lines). If value of $b$ is increased the curve goes downward direction and if value of $b$ is decreased the curve goes upward direction. But the nature of the curve is similar. In fig.11(a-b), we present our results of $x$-evolution of gluon structure function for $K_1(x) = ax^b$ in relation $b = a^y$, for $y$ minimum ( lower thick solid lines) and maximum (upper thick solid lines) at same parameter values $a$ = 372, $b$ = 4 in fig.11(a) and $a$ = 135, $b$ = 1.8 in fig.11(b) and for representative values of $Q^2$ given in each figure, and compare them with recent global parameterization [37]. We observed that result of $x$-evolution of gluon structure function in relation $b = a^y$, for $y$ maximum (long dashed lines) coincide with result of $x$-evolution of gluon structure function for $y$ minimum (lower thick solid lines) when $a$ = 375, $b$ = 4.7 in fig.11(a) and $a$ = 134, $b$ = 2 in fig.11(b). That means if $y$ varies from minimum to maximum, then value of parameter $a$ varies from 372 to 375 and $b$ varies from 4 to 4.7 in fig. 11(a) and $a$ varies from 135 to 134 and $b$ varies from 1.8 to 2 in fig.11(b).

In fig.12(a-b), we present our results of $x$-distribution of gluon structure functions for $K_1(x) = ce^{-dx}$ , where $c$ and $d$ are constants for representative values of $Q^2$ given in each figure, and compare them with recent global parameterization [37] in the relation $b = a^y$ for $y$ minimum ( thick solid lines). In fig.12(a), we observed that agreement of the results with parameterization is found to be very poor for any values of $c$ and $d$ at small-$x$ and agreement is found to be good at high-$x$ at $c = 300$ and $d = -3.8$. In fig.12(b) agreement of the results with parameterizations is found to be good at $c = 5$ and $d = -28$. In the same figures we present the sensitivity of our results for different values of $c$ at fixed value $d$ (dashed lines). We observe that if value of $c$ is increased, the curve goes upward direction and if value of $c$ is decreased, the curve goes downward direction. But the nature of the curve is similar. In fig.13(a-b), we present the sensitivity of our results of gluon structure functions for different values of $d$ at fixed value of $c$ (dashed lines). If value of $d$ is increased the curve goes downward direction and if value of $d$ is decreased the curve goes upward direction. But the nature of the curve is similar. In fig.14(a-b), we present our results of $x$-evolution of gluon structure function for $K_1(x) = ce^{-dx}$ in relation $b = a^y$, for $y$ minimum ( lower thick solid lines) and maximum (upper thick solid lines) at same parameter values $c$ = 300, $d$ = -3.8 in fig.14(a) and $c$ = 5, $d$ = -28 in fig.14(b) and for representative values of $Q^2$ given in each figure, and compare them with recent global parameterization [37]. We observed that result of $x$-evolution of gluon structure function in relation $b = a^y$, for $y$ maximum (long dashed lines) coincide with result of $x$-evolution of gluon structure function for $y$ minimum (lower thick solid lines) when $c$ = 300, $b$ = -3.6 in fig.14(a) and $c$ = 5, $d$ = -25.3 in fig.14(b). That means if $y$ varies from minimum to maximum, then value of parameter $d$ varies from -3.8 to -3.6 in fig.14(a) and $d$ varies from -28 to -25.3 in fig.14(b). In these cases values of parameter $c$ remain constant.

In fig.15(a-b), we present our results of $x$-distribution of light sea quark structure functions for $K(x) = k$ (constant) for representative values of $Q^2$ = 10 GeV$^2$ (fig.15(a)) and $Q^2$ = 10$^4$ GeV$^2$ (fig.15(b)) and compare them with recent global parameterizations (thin solid lines) [37] in the relation $b = a^y$ for $y$ =2 (thick solid lines). Since our theory is in small-$x$ region and does not explain the peak portion for $u$ & $d$, so in each the data point for $x$-value just below 0.1 for $s$ and 0.01 for $u$ & $d$ has been taken as input to test the evolution equation. We observed that agreement of the results (thick solid line) with parameterization is good in the small-$x$ region at $k$ = 60, 590 for $u$ & $d$ and $k$ = 210, 520 for $s$ in fig.15(a) and fig.15(b) respectively. In the same figures we present the sensitivity of our results (dashed lines) for different constant values of $k$. We observe that if value of $k$ is increased or decreased, the curve goes upward or downward direction respectively. But the nature of the curves is similar.

In figs.16(a-b) and 17(a-b), we present our results of $x$-distribution of light sea quark structure functions for $K(x) = ax^b$, where $a$ and $b$ are constants for representative values of $Q^2$ = 10 GeV$^2$ (figure 16(a-b)) and $Q^2$ = 10$^4$ GeV$^2$ (fig.17(a-b)) and compare them with recent global parameterizations (thin solid lines) [37] in the relation $b = a^y$ for $y$ = 2 (thick solid lines). In each the data point for $x$-value just below 0.1 for s and 0.01 for $u$ & $d$ has been taken as input to test the evolution equation for the reason explained above. We observed that agreement of the results (thick solid line) with parameterization is found to be good in the small-$x$ region at $a$ = 135 & $b$ = 0.33 for $u$ & $d$ and $a$ = 130 & $b$ = 0.35 for $s$ at $Q^2$ = 10 GeV$^2$ in fig.16(a-b) and $a$ = 211 & $b$ = 0.25 for $u$ & $d$ and $a$ = 260 & $b$ = 0.29 for s at $Q^2$ = 10$^4$ GeV$^2$ in fig.17(a-b). In the same figures we present the sensitivity of our results (dashed lines) for different values of $a$ and $b$. We observe that if value of $a$ is increased or decreased, the curve goes upward or downward direction, and if value of $b$ is increased or decreased the curve goes downward or upward direction. But the nature of the curves is similar. Figs.16(a) and 17(a) give the sensitivity of $a$ and figs. 16(b) and 17(b) give the sensitivity of $b$.

In figs.18(a-b) and 19(a-b) we present our results of $x$-distribution of light sea quark structure functions for $K(x) = ce^{-dx}$, where $c$ and $d$ are constants for representative values of $Q^2$ = 10 GeV$^2$ (fig.18(a-b)) and $Q^2$ = 10$^4$ GeV$^2$ (fig.19(a-b)) and compare them with recent global parameterizations (thin solid lines) [37] in the relation $b = a^y$ for $y$ = 2 (thick solid lines). Inputs have been taken as before. We observed that agreement of the results (thick solid line) with parameterization is found to be good in the small-$x$ region at $c$ = 47.8 & $d$ = -1 for $u$, $d$ and $c$ = 32.5 & $d$ = -20 for $s$ at $Q^2$ = 10 GeV$^2$ in figs.18(a-b) and $c$ = 465 & $d$ = -.4 for $u$ ,$d$ and $c$ = 385 & $d$ = -25 for $s$ at $Q^2 = 10^4$ GeV$^2$ in figs.19(a-b). In the same figures we present the sensitivity of our results (by dashed lines) for different values of $c$ and $d$. We observe that if value of $c$ or $d$ is increased or decreased, the curve goes upward or downward direction. But the nature of the curves is similar. Figs.18(a) and 19(a) give the sensitivity of $c$ and figs.18(b) and 19(b) give the sensitivity of $d$. We observed that for $x$-evolutions of light sea quark structure functions, results for $y$ minimum and maximum in $\beta = \alpha^y$ relation has not any significant difference.

## 4. Conclusion

We solve DGLAP evolution equation in LO and NLO using Taylor expansion method and derive $t$ and $x$-evolutions of various structure functions and compare them with global data and parameterizations with satisfactory phenomenological success. It has been observed that though we have derived a unique $t$-evolution for deuteron, proton, neutron, difference and ratio of proton and neutron structure functions in LO and NLO, yet we can not establish a completely unique $x$-evolution for deuteron structure function in LO and NLO due to the relation $K(x)$ between singlet and gluon structure functions. $K(x)$ may be in the forms of a constant, an exponential function or a power function and they can equally produce required $x$-distribution of deuteron structure functions. But unlike many parameter arbitrary input $x$-distribution functions generally used in the literature, our method requires only one or two such parameters. On the other hand, we observed that the Taylor expansion method is mathematically simpler in comparison with other methods available in the literature. Explicit form of $K(x)$ can actually be obtained only by solving coupled DGLAP evolution equations for singlet and gluon structure functions. From the preliminary work, we see that the same method can also be used to solve evolution equations for spin dependent structure functions too. Though we study LO evolution equation for spin structure function, we hope that it can be extendable to NLO also. So we see that this simple method may have a wide application in solving quark and gluon evolution equations.

## Figure Captions

**Fig.1(a-d):** Results of $t$-evolutions of deuteron, proton, neutron and difference of proton and neutron structure functions (dashed lines for LO and solid lines for NLO) for the representative values of $x$ in LO and NLO for NMC data. For convenience, value of each data point is increased by adding $0.2i$, where $i$ = 0, 1, 2, 3 ... are the numberings of curves counting from the bottom of the lowermost curve as the 0-th order. Data points at lowest-$Q^2$ values in the figures are taken as input.

**Fig.2:** Results of $t$-evolutions of proton structure functions $F_2^p$ (dashed lines for LO and solid lines for NLO) with recent global paramatrization (long dashed lines) for the representative values of $x$ given in the figures. Data points at lowest-$Q^2$ values in the figures are taken as input. For convenience, value of each data point is increased by adding $0.5i$, where $i$ = 0, 1, 2, 3, ... are the numberings of curves counting from the bottom of the lowermost curve as the 0- th order.

**Fig.3(a-e):** Results of $t$-evolutions of gluon, light sea and valence quark structure functions qualitatively for the representative values of $x$ given in the figures for $y$ = 2 (upper solid lines and upper dashed lines in fig.3(a), solid lines in fig.3(b-e)) and y maximum (lower solid lines and lower dashed lines in fig.3(a), dashed lines in fig.3(b-e)) in $\beta = \alpha^y$ relation. We have taken arbitrary inputs from recent global parameterizations MRST2001 (for solid lines in fig.3(a)) and MRST2001J (for dashed lines in fig.3(a)) and MRS data (in fig.3(b)) at $Q_0^2$= 1 GeV$^2$ and $Q_0^2$

= 4 GeV$^2$ . For convenience, value of each data point is increased by adding 5 and 2 for $x$ = 0.1 and $x$ = .05 respectively in fig.3(a) and decreased by subtracting 1 for $x$ = 0.1 in fig.3(b) and increased by adding 4, 5, 6, 7 for $x$ = 0.0001, 0.001, 0.01, 0.1 respectively in fig.3(c).

**Fig.4(a-b):** Results of $x$-distribution of deuteron structure functions $F_2^d$ in LO for $K(x) = k$ (constant) (solid lines), $K(x) = ax^b$ (dashed lines) and for $K(x) = ce^{-dx}$ (dotted lines), where $k$ = 4.5, $a$ = 4.5, $b$ = 0.01, $c$ = 5, $b$ = 1 and in NLO for $K(x) = ax^b$ (solid lines), and for $K(x) = ce^{-dx}$ (dotted lines), where $a$ = 5.5, $b$ = 0.016, $c$ = 0.28, and $d$ = -3.8 and for representative values of $Q^2$ given in each figure, and compare them with NMC deuteron low-$x$ low-$Q^2$ data. In each the data point for $x$-value just below 0.1 has been taken as input $F_2^d(x_0, t)$. For convenience, value of each data point is increased by adding 0.2$i$, where $i$ = 0, 1, 2, 3, ... are the numberings of curves counting from the bottom of the lowermost curve as the 0-th order.

**Fig.5:** Sensitivity of our results of $x$-distribution of deuteron structure function in the relation b = a$^y$ for $y$ minimum for different values of $k$, $a$, $b$, $c$ and $d$ in LO.

**Fig.6:** Sensitivity of our results of $x$-distribution of deuteron structure function in the relation b = a$^y$ for $y$ minimum for different values of $a$, $b$, $c$ and $d$ in NLO.

**Fig.7:** Sensitivity of our results of $x$-distribution of deuteron structure function for different values of $T_0$ at best fit of $K(x) = ax^b$ and $K(x) = ce^{-dx}$ in the relation b = a$^y$ for $y$ minimum.

**Fig.8:** $T(t)^2$ and $T_0T(t)$, where $T(t)$ = $á_s(t)/2ð$ against $Q^2$ in the $Q^2$ range $0 \leq Q^2 \leq 30$ GeV$^2$.

**Fig.9(a-b):** Results of $x$-distribution of gluon structure functions for $K(x) = ax^b$ , where $a$ and $b$ are constants for representative values of $Q^2$ given in each figure, and compare them with recent global parameterization in the relation b = a$^y$ for $y$ minimum (upper thick solid lines). In the same figures we present the sensitivity of our results for different values of $a$ at fixed value $b$. Here we take $b$ = 4 in fig.9(a) and $b$ = 1.8 in fig.9(b).

**Fig.10(a-b):** Sensitivity of our results for different values of $b$ at fixed value of $a$. Here we take $a$ = 372 in fig.10(a) and $a$ = 135 in fig.10(b).

**Fig.11(a-b):** Results of $x$-evolution of gluon structure function for $K(x) = ax^b$ in relation b = a$^y$, for $y$ minimum (lower thick solid lines) and maximum (upper thick solid lines) at $a$ = 372, $b$ = 4 in fig.11(a) and $a$ = 135, $b$ = 1.8 in fig.11(b) and for representative values of $Q^2$ given in each figure, and compare them with recent global parameterization. We also plot result of $x$-evolution of gluon structure function in relation $â=á^y$, for $y$ maximum (long dashed lines) at $a$ = 375, $b$ = 4.7 in fig.11(a) and $a$ = 134, $b$ = 2 in fig.11(b).

**Fig.12(a-b):** Results of $x$-distribution of gluon structure functions for $K(x) = ce^{-dx}$ , where $c$ and $d$ are constants for representative values of $Q^2$ given in each figure, and compare them with recent global parameterization in the relation b = a$^y$ for $y$ minimum ( thick solid lines). In the same figures we present the sensitivity of our results for different values of $c$ at fixed value $d$. Here we take $d$ = -3.8 in fig.12(a) and $d$ = -288 in fig.12(b).

**Fig.13(a-b):** Sensitivity of our results for different values of $d$ at fixed value of $c$. Here we take $c$ = 300 in fig.13(a) and $c$ = 5 in fig.13(b).

**Fig.14(a-b):** Results of $x$-evolution of gluon structure function for $K(x) = ce^{-dx}$ in relation b = a$^y$, for $y$ minimum (lower thick solid lines) and maximum (upper thick solid lines) at $c$ = 300, $d$ = -3.8 in fig.14(a) and $c$ = 5, $d$ = -28 in fig.14(b) and for representative values of $Q^2$ given in each figure, and compare them with recent global

parameterization. We plot the results of x-evolution of gluon structure function in relation $b = a^y$, for *y* maximum (long dashed lines) at *c* = 300, *b* = -3.6 in fig.14(a) and *c* = 5, *d* = -25.3 in fig.14(b).

**Fig.15(a-b):** Results of *x*-distribution of light sea quark structure functions for *K*(*x*) = *k* (constant) for representative values of $Q^2$ given in each figure, and compare them with recent global parameterization (thin solid lines) in the relation $b = a^y$ for *y* minimum (thick solid lines). In the same figures we present the sensitivity of our results (dashed lines) for different constant values of *K*(*x*).

**Fig.16(a-b):** Results of *x*-distribution of light sea quark structure functions for $K(x) = ax^b$, where *a* and *b* are constants for $Q^2$ = 10 $GeV^2$ and compare them with recent global parameterization (thin solid lines) in the relation $b = a^y$ for y minimum (thick solid lines). In the same figures we present the sensitivity of our results (dashed lines) for different values of *a* and *b*.

**Fig.17(a-b):** Results of *x*-distribution of light sea quark structure functions for $K(x) = ax^b$, where *a* and *b* are constants for $Q^2 = 10^4$ $GeV^2$ and compare them with recent global parameterization (thin solid lines) in the relation $b = a^y$ for *y* minimum (thick solid lines). In the same figures we present the sensitivity of our results (dashed lines) for different values of *a* and *b*.

**Fig.18(a-b):** Results of *x*-distribution of light sea quark structure functions for $K(x) = ce^{-dx}$, where *c* and *d* are constants for $Q^2$ = 10 $GeV^2$, and compare them with recent global parameterization (thin solid lines) in the relation $b = a^y$ for *y* minimum (thick solid lines). In the same figures we present the sensitivity of our results (dashed lines) for different values of *c* and *d*.

**Fig.19(a-b):** Results of *x*-distribution of light sea quark structure functions for $K(x) = ce^{-dx}$, where *c* and *d* are constants for $Q^2 = 10^4$ $GeV^2$, and compare them with recent global parameterization (thin solid lines) in the relation $b = a^y$ for *y* minimum (thick solid lines). In the same figures we present the sensitivity of our results (dashed lines) for different values of *c* and *d*.

## Acknowledgement

One of us (JKS) is grateful to UGC, New Delhi for the financial assistance to this work in the form of a major research project.

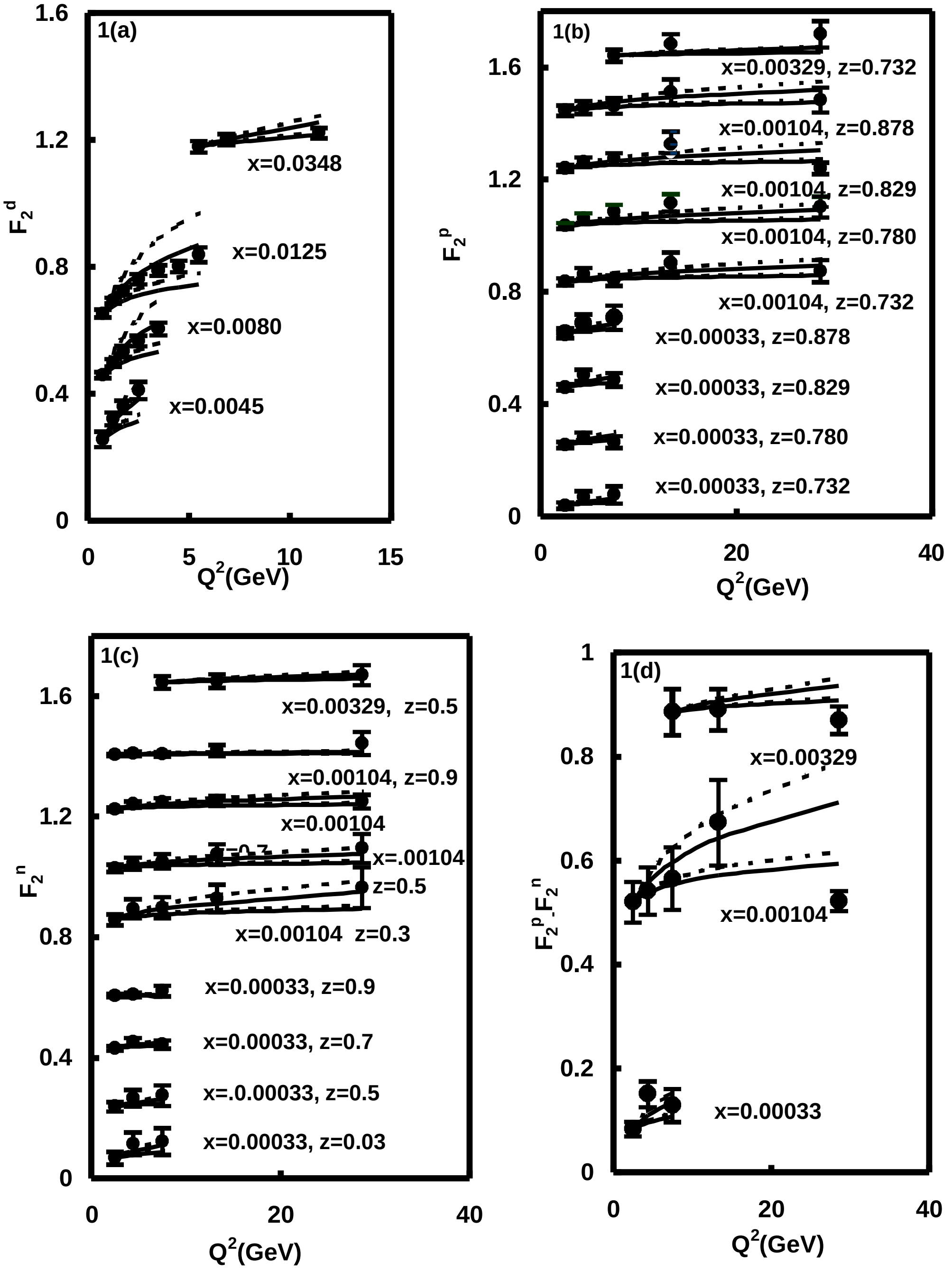

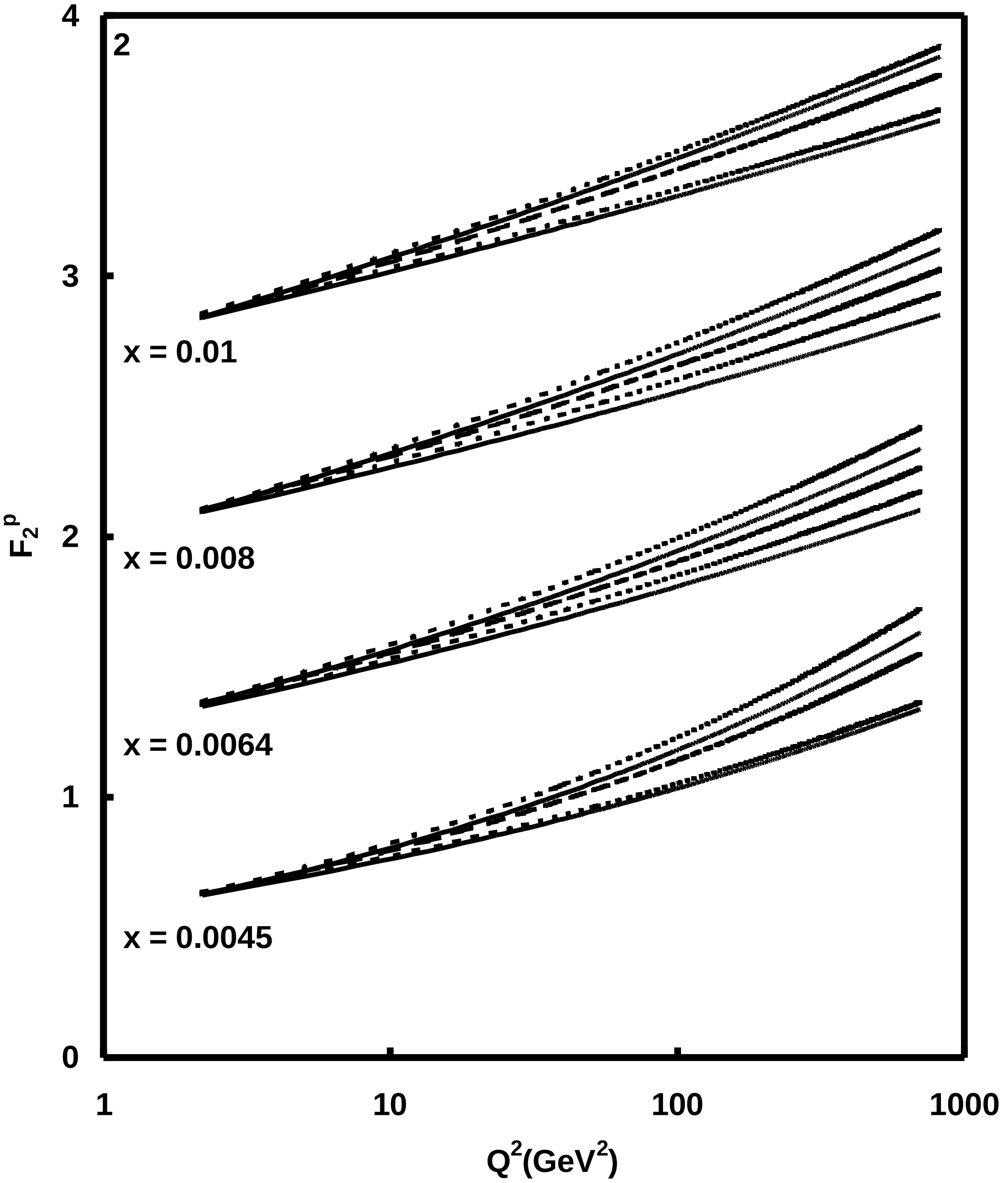

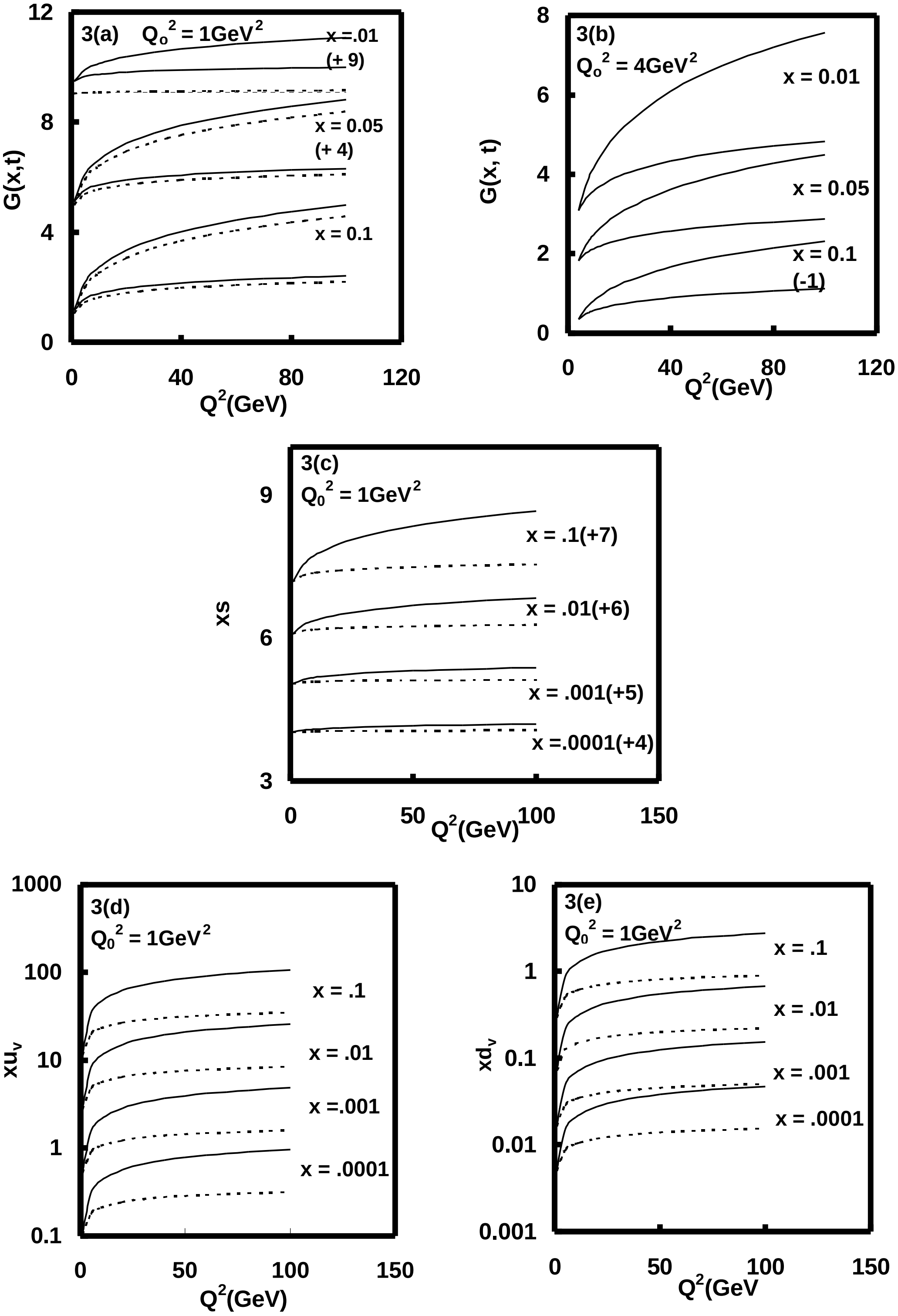

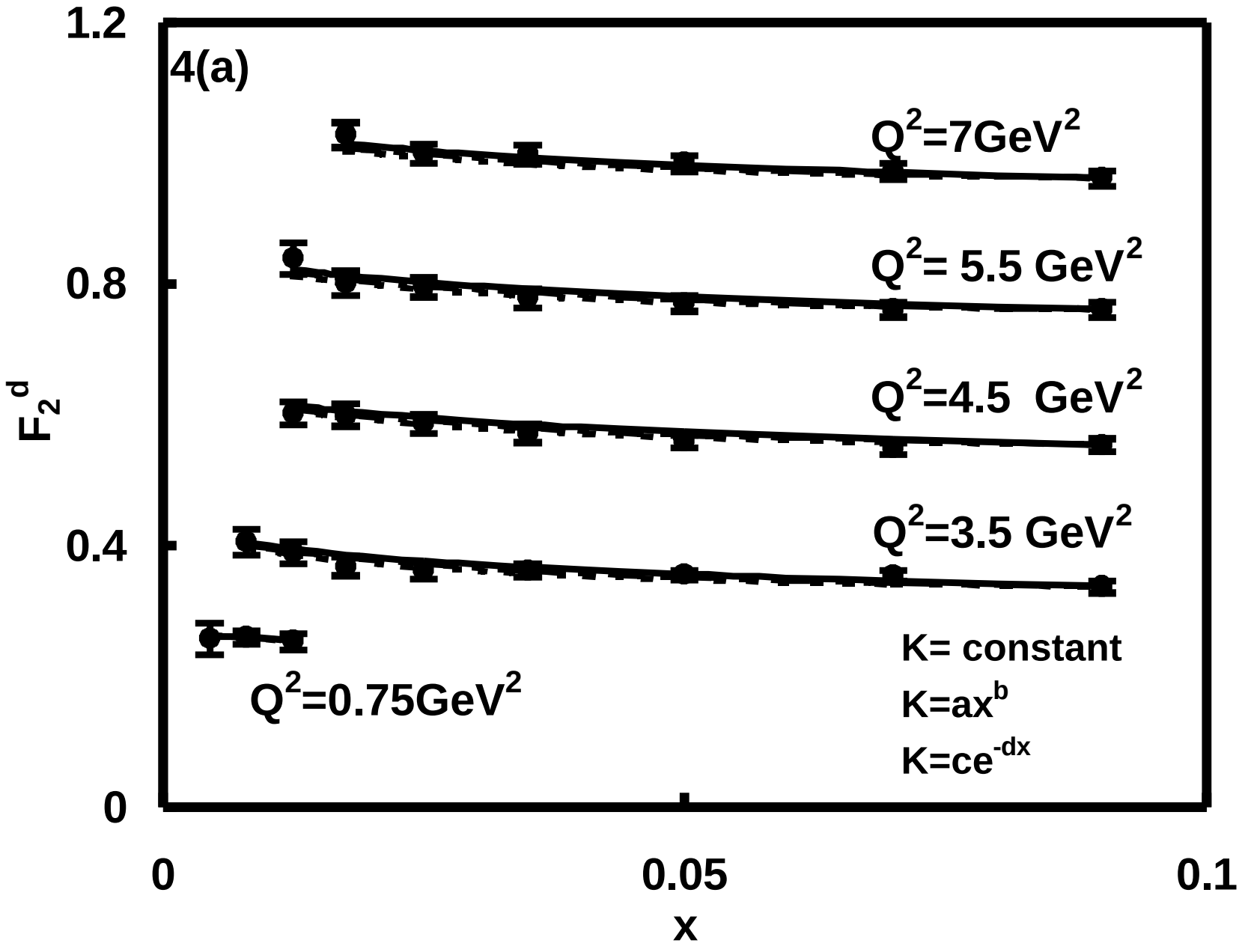

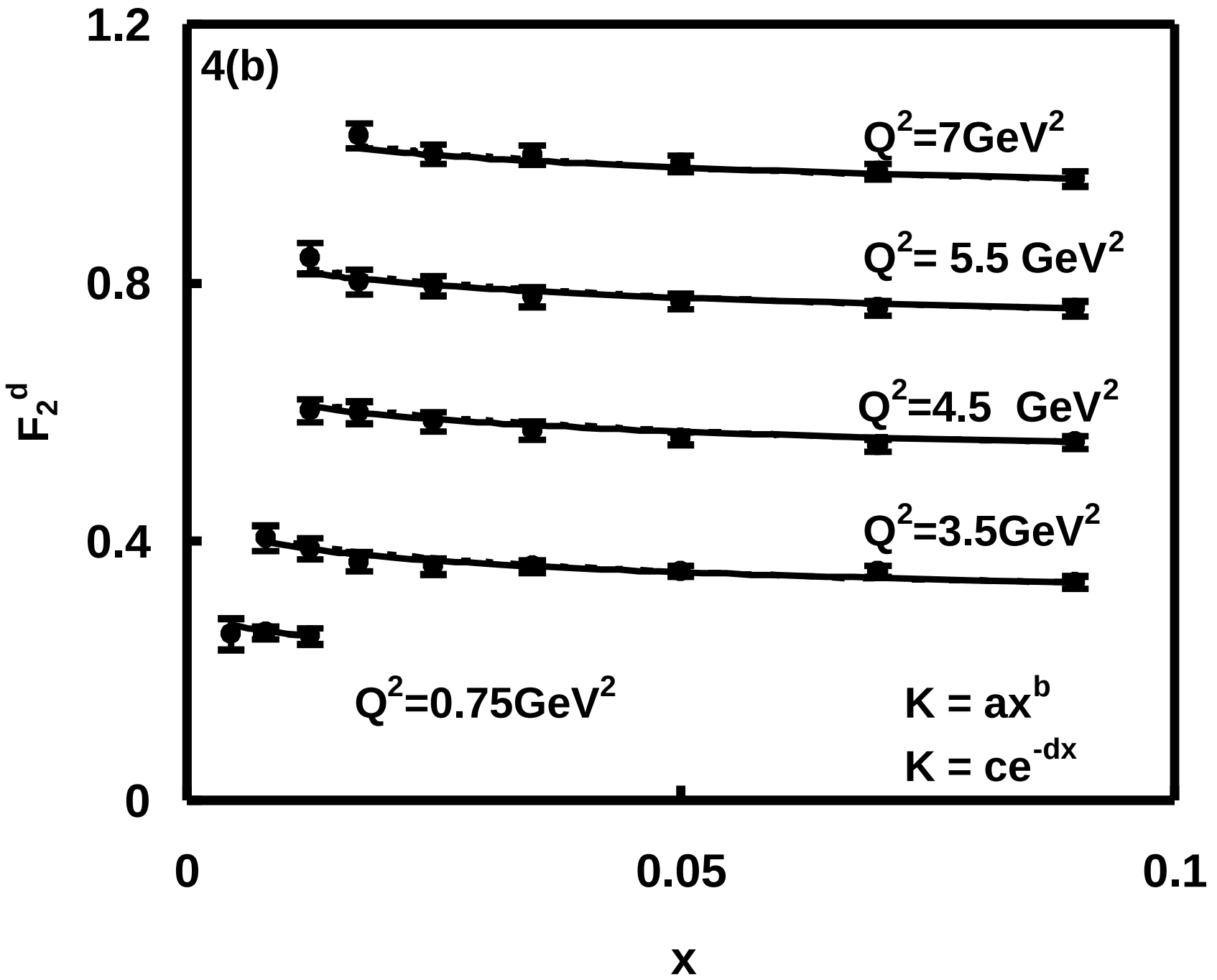

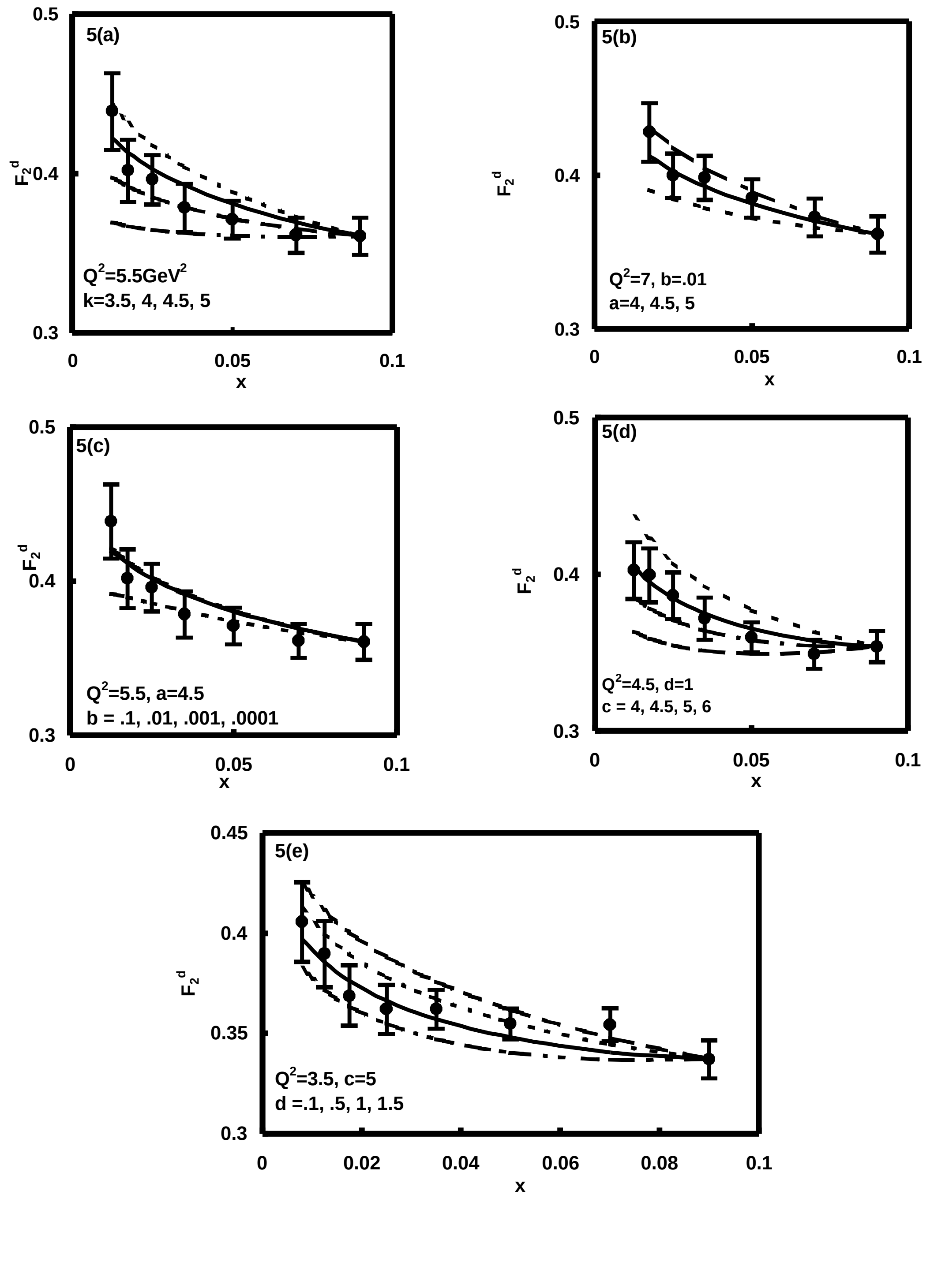

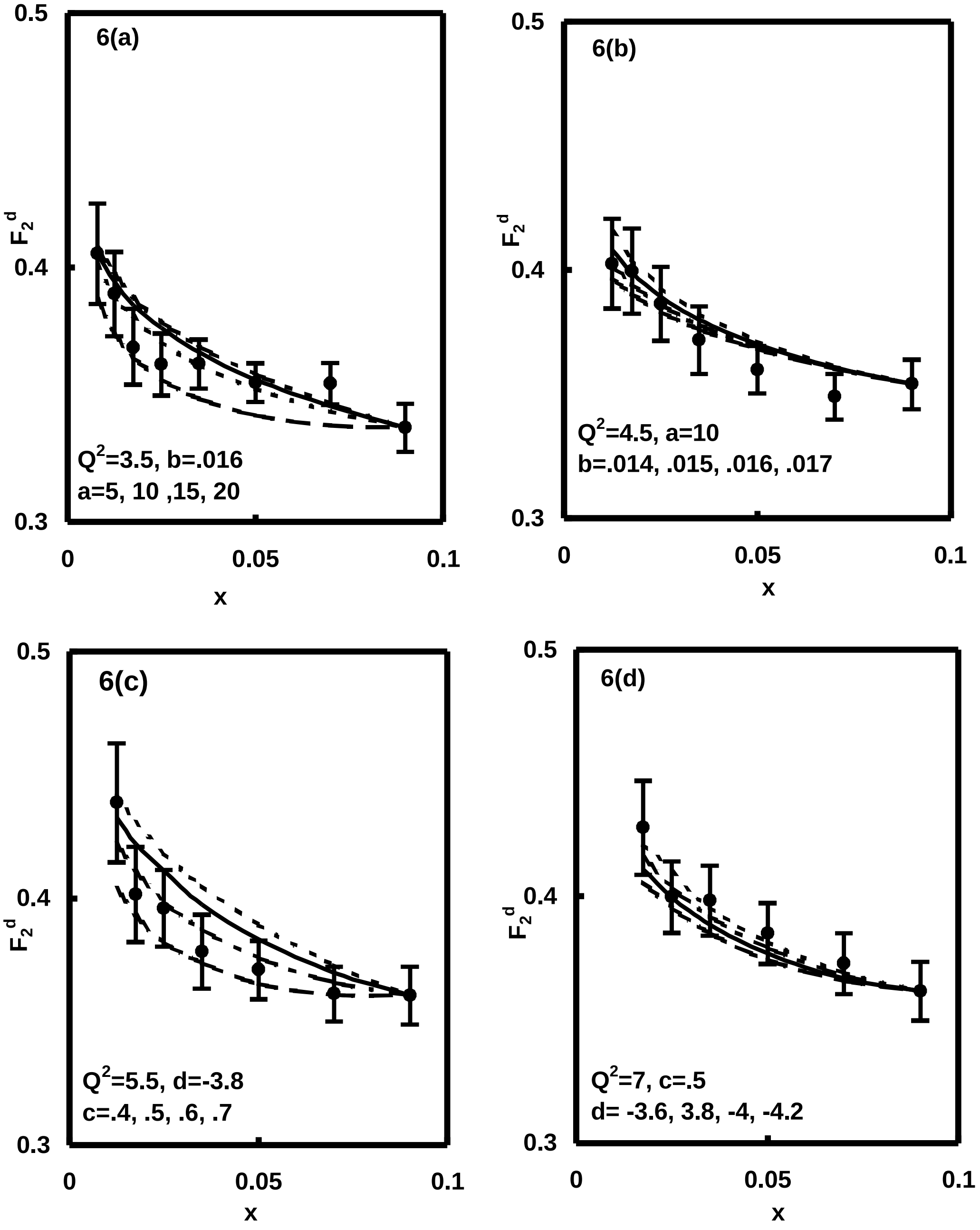

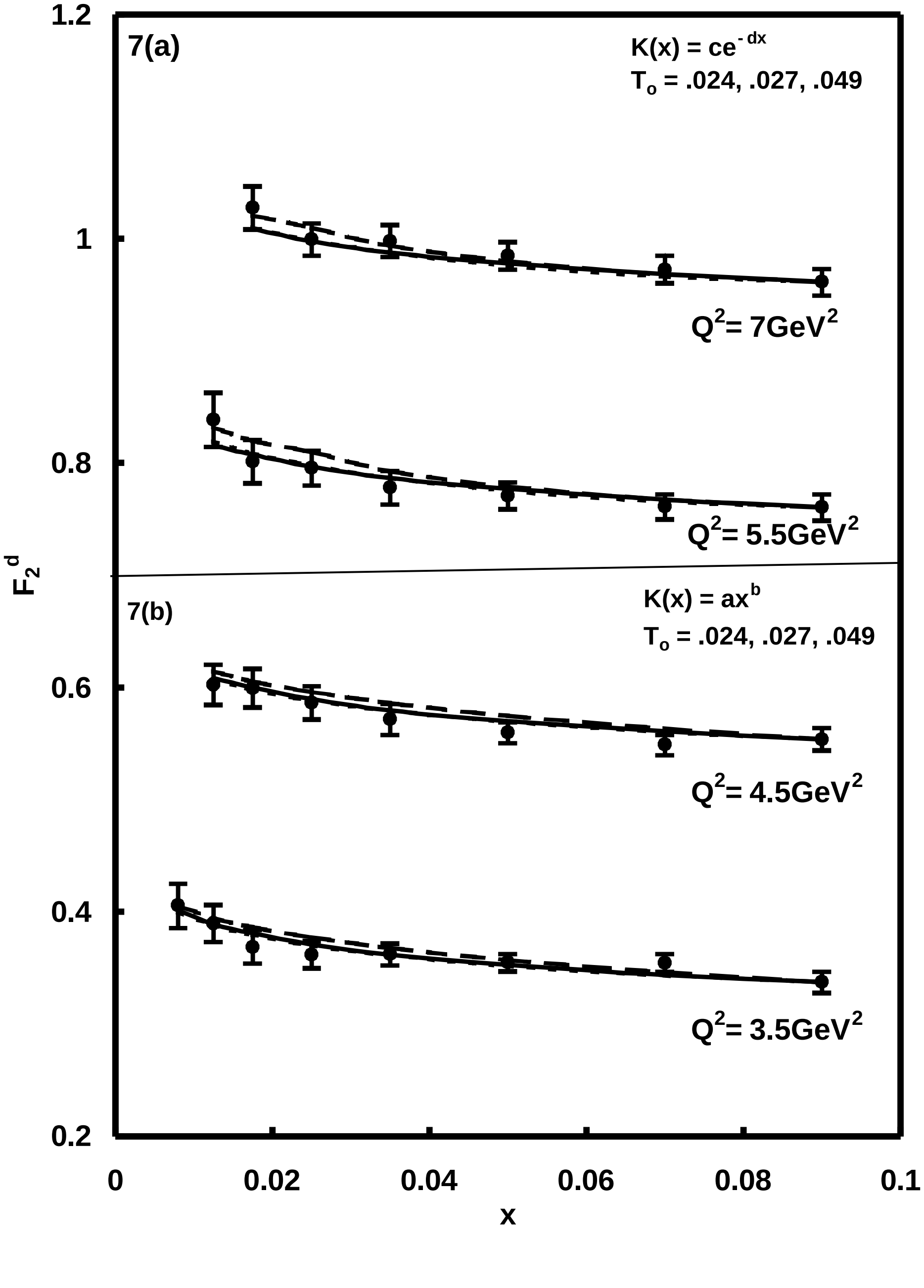

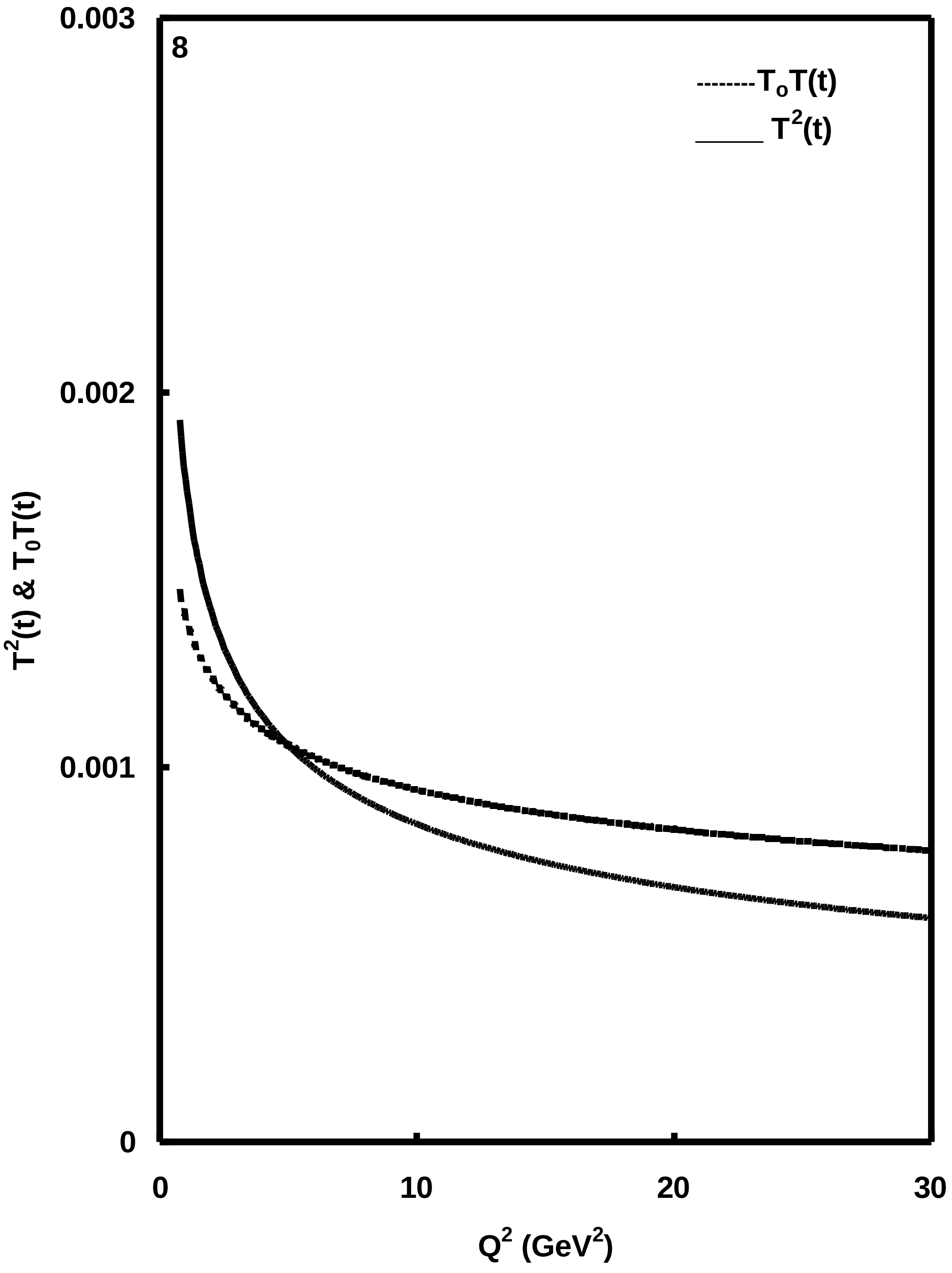

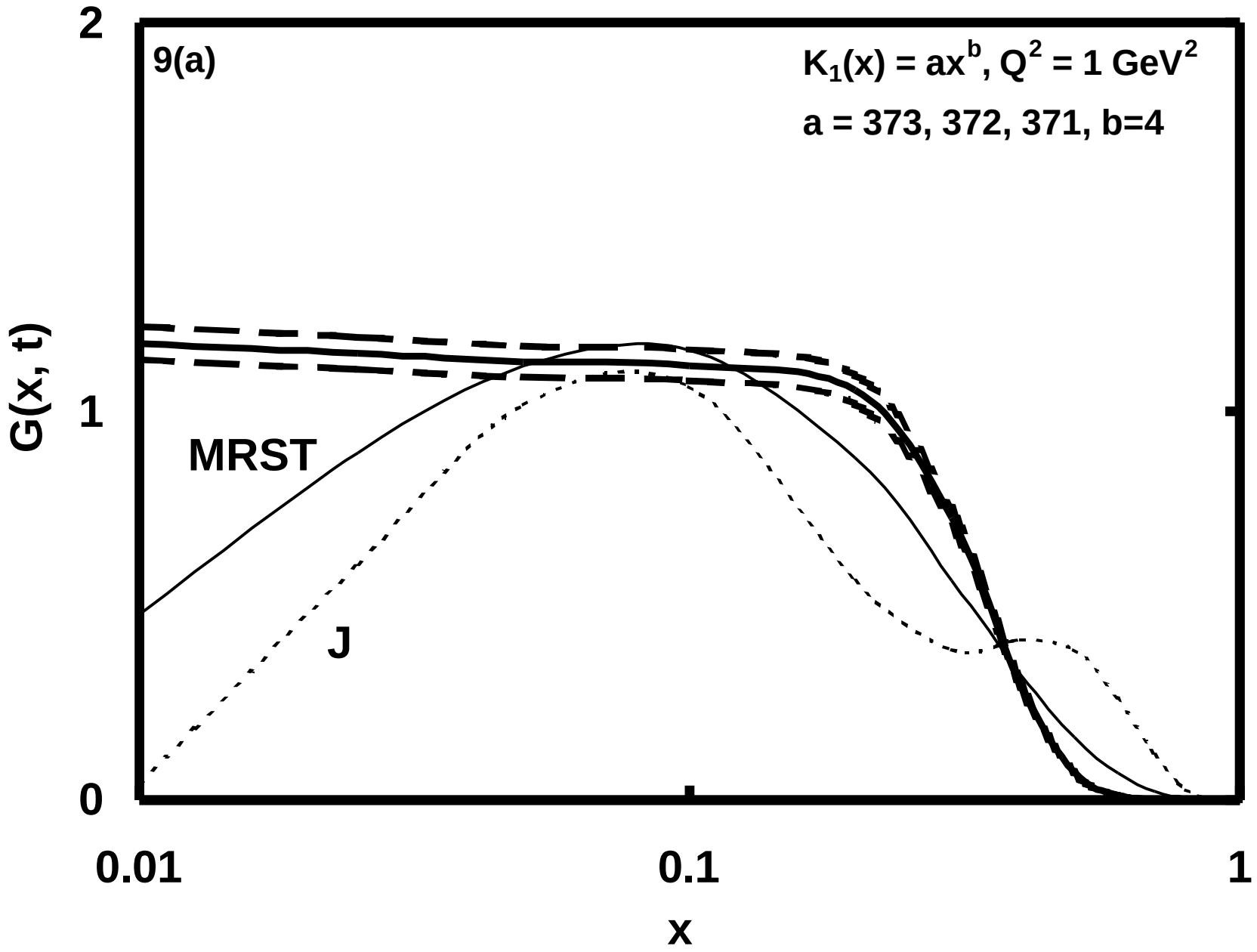

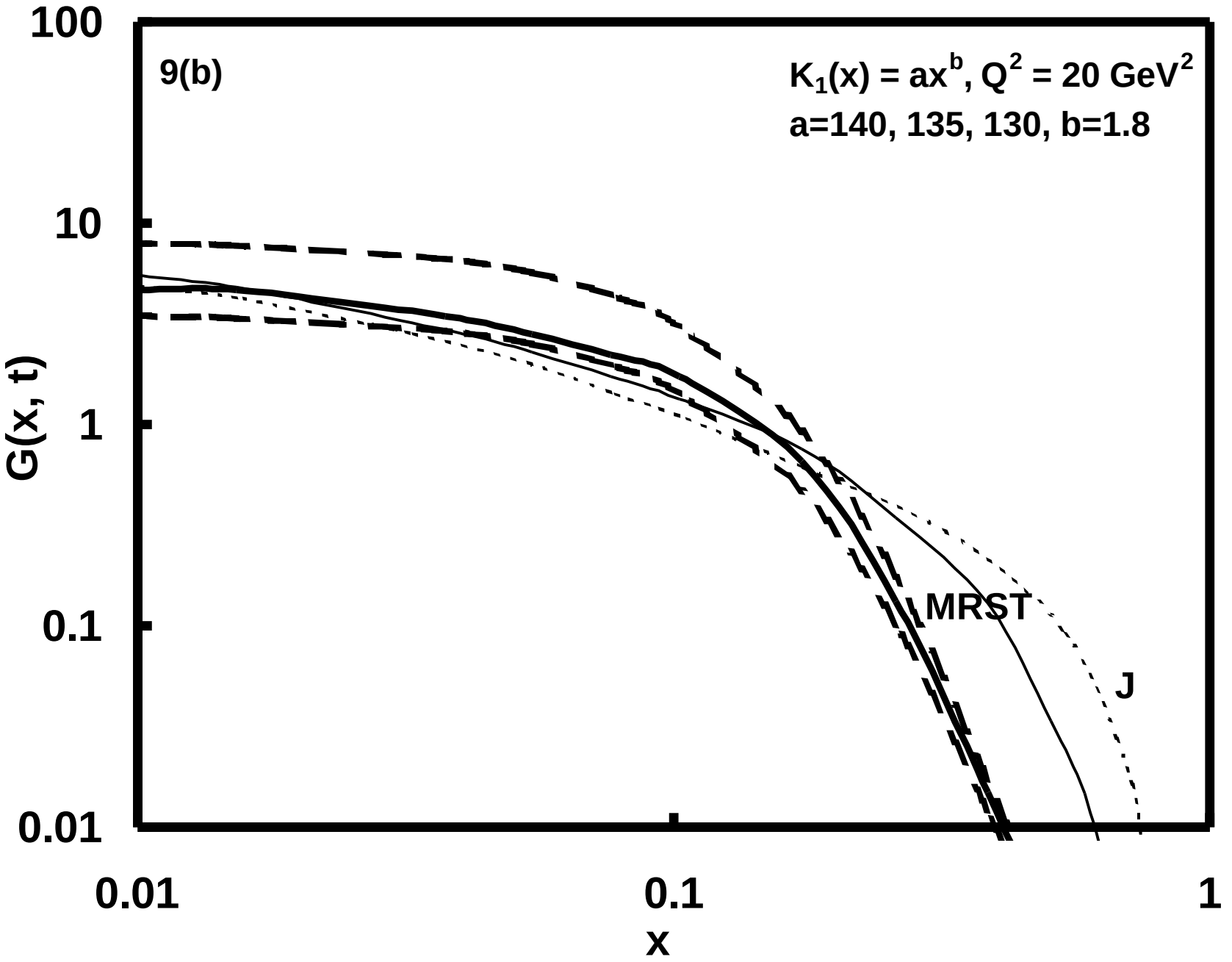

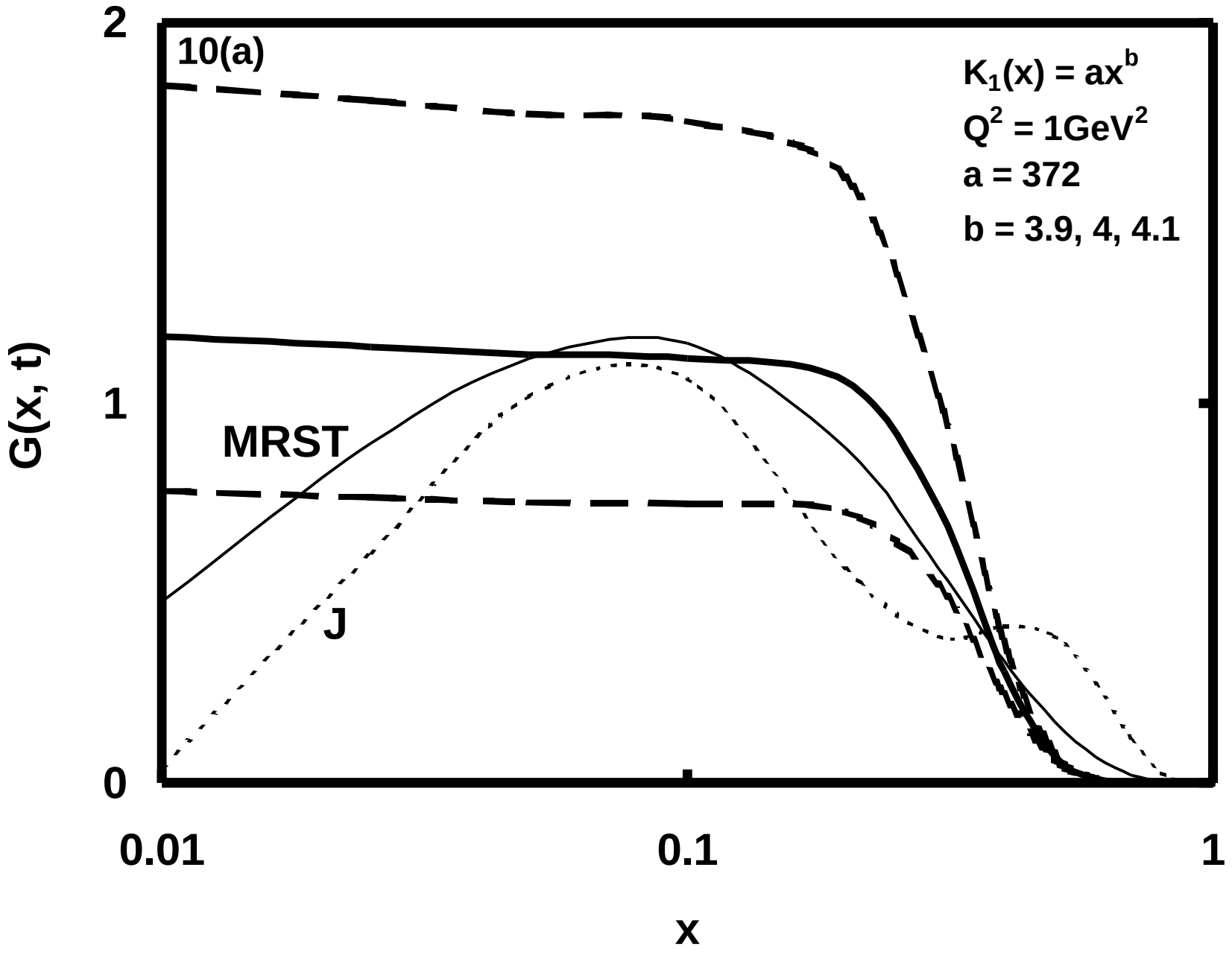

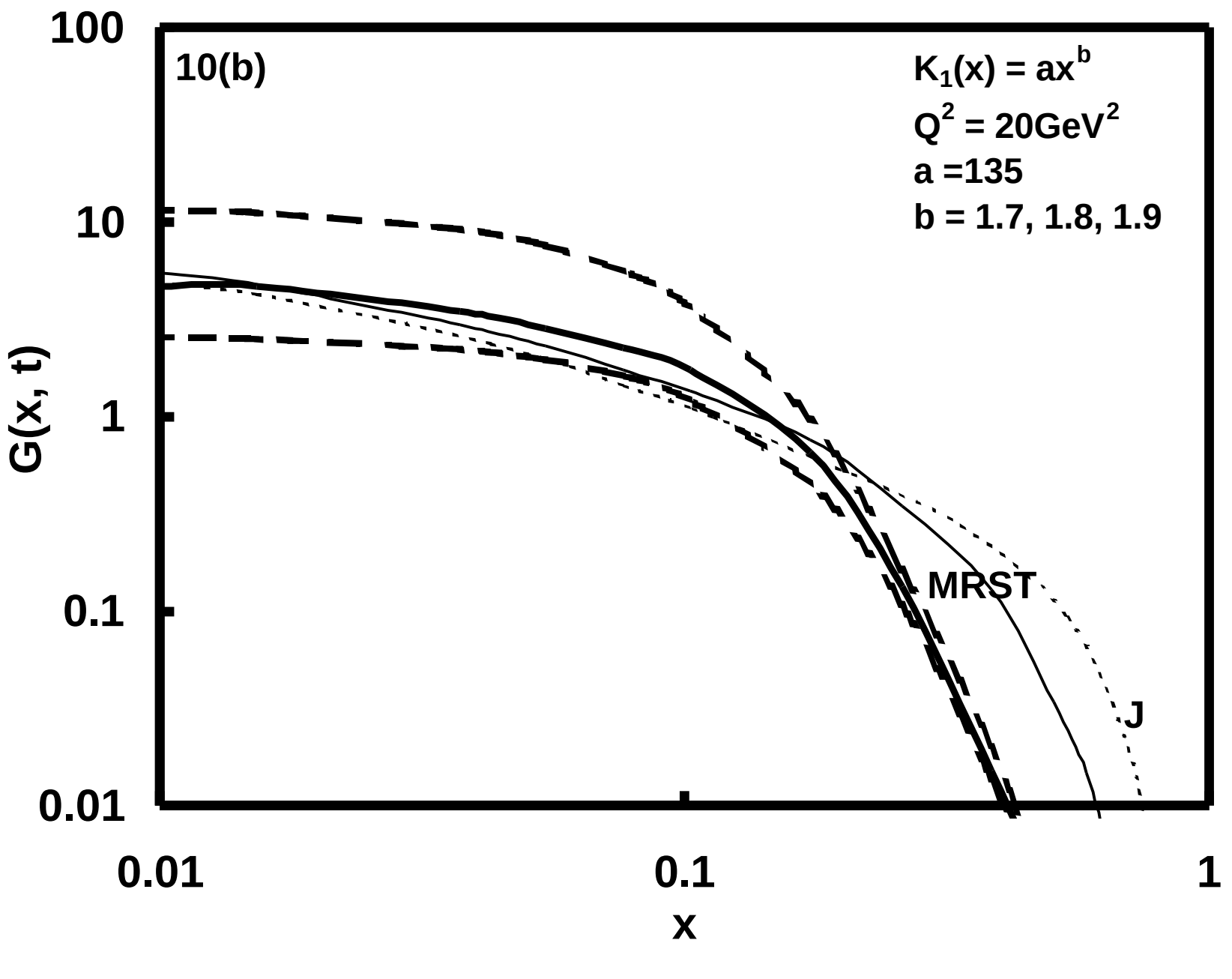

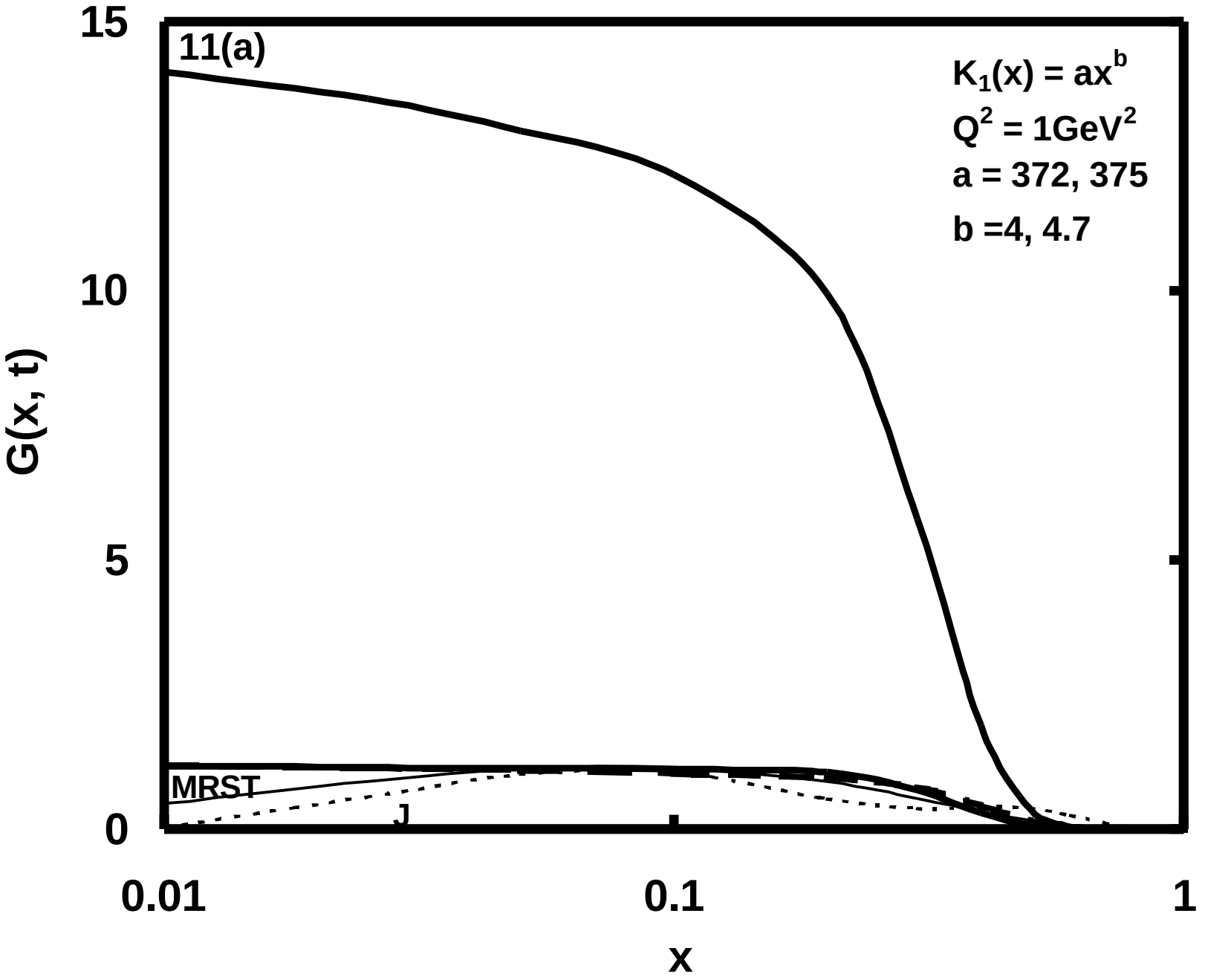

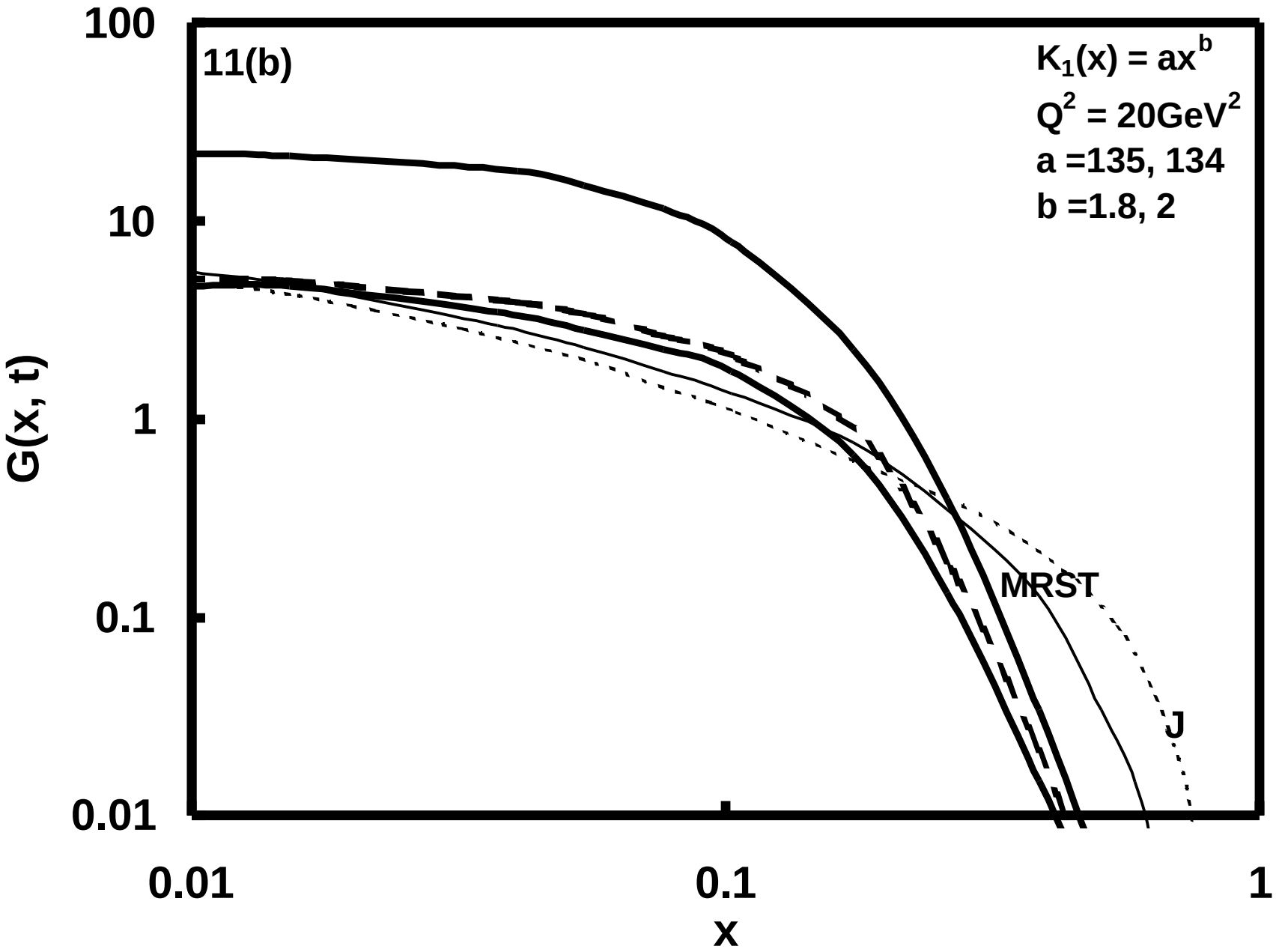

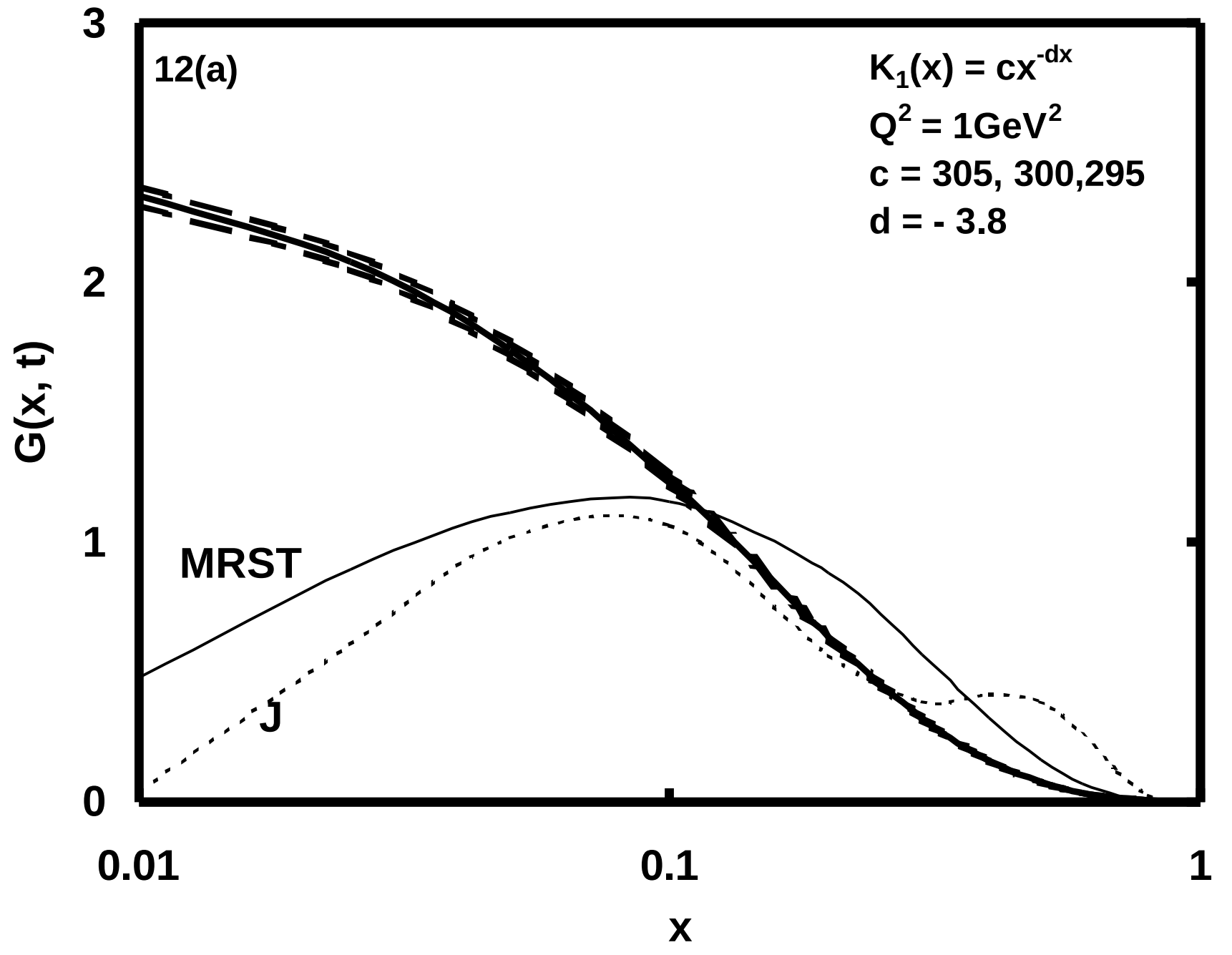

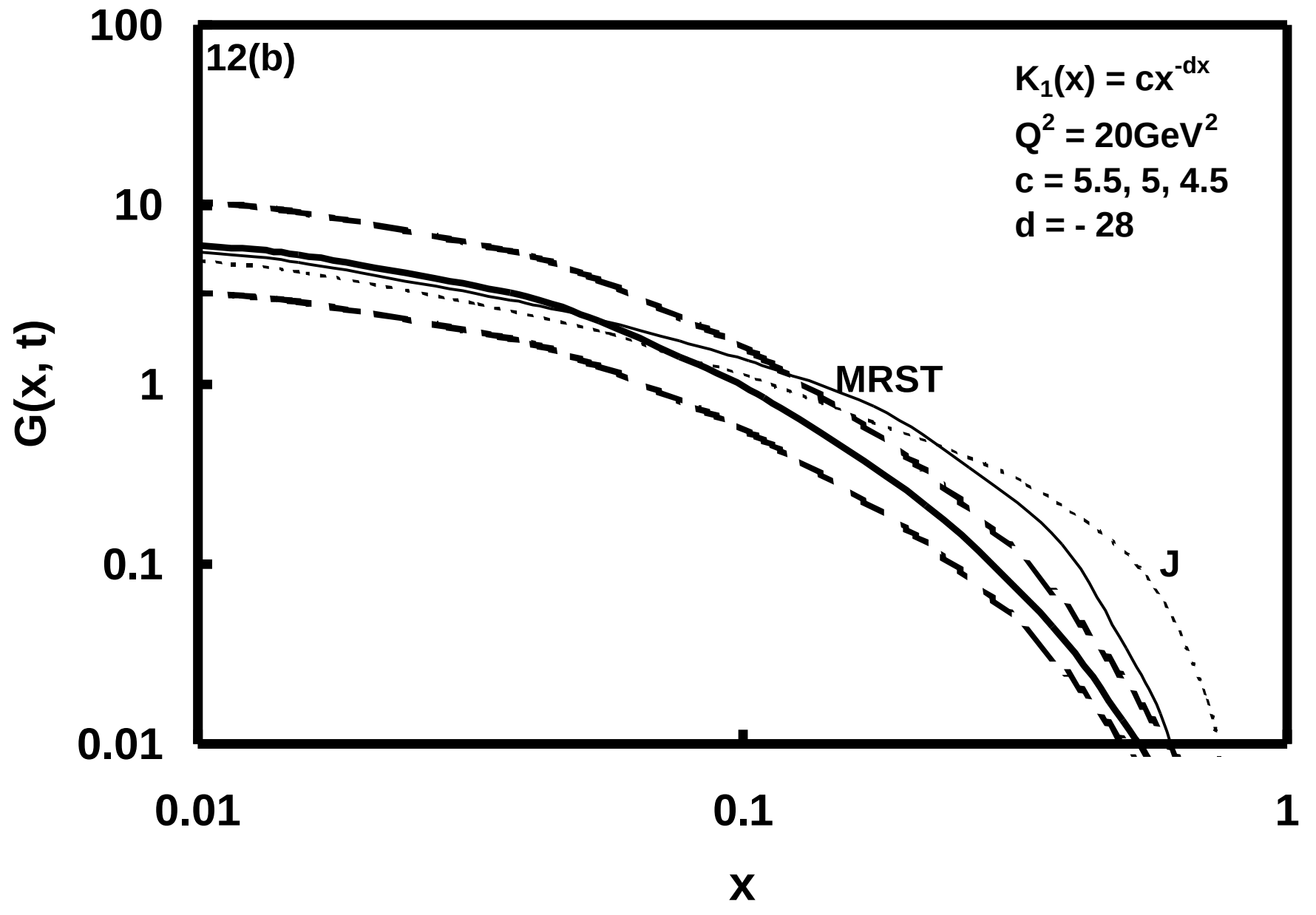

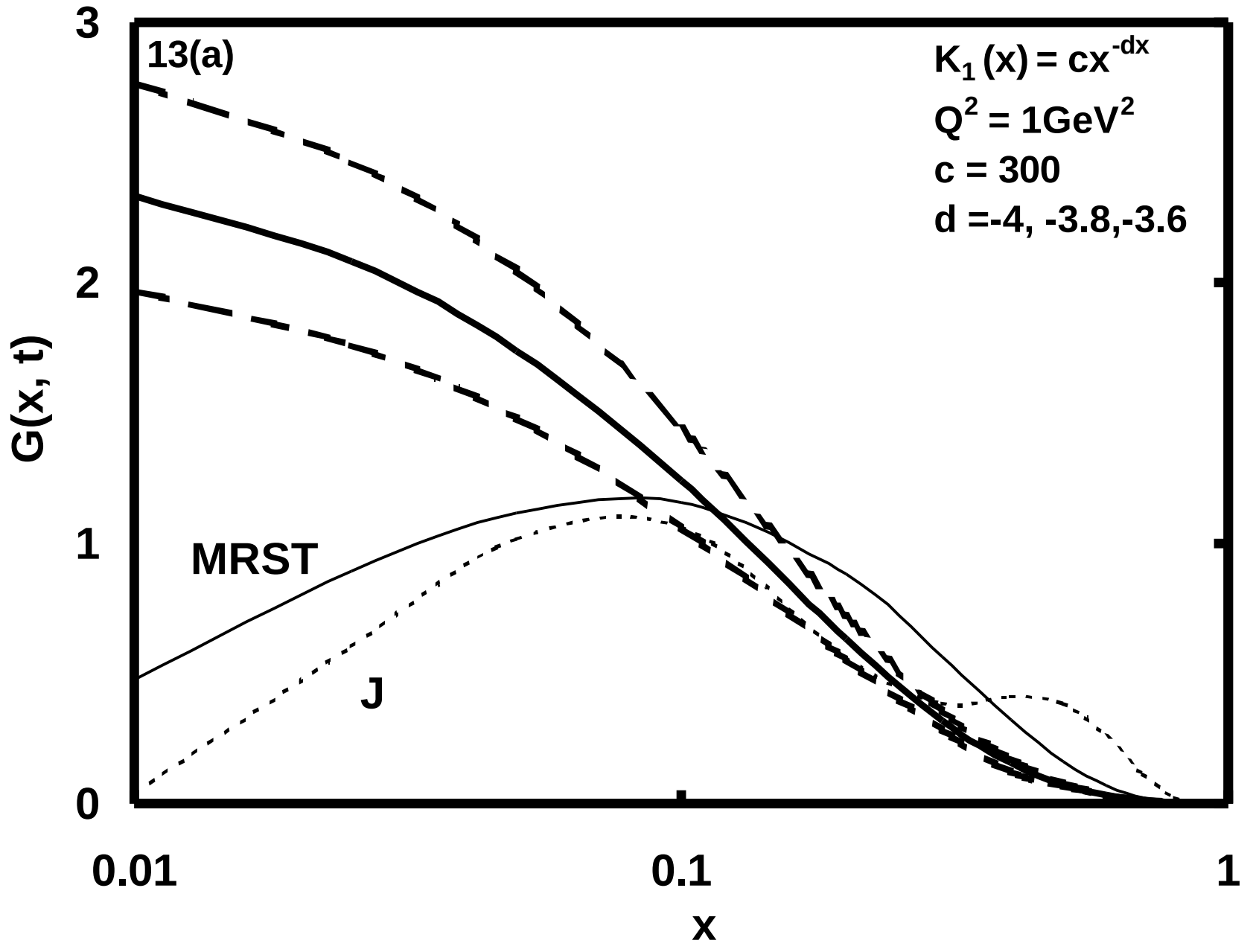

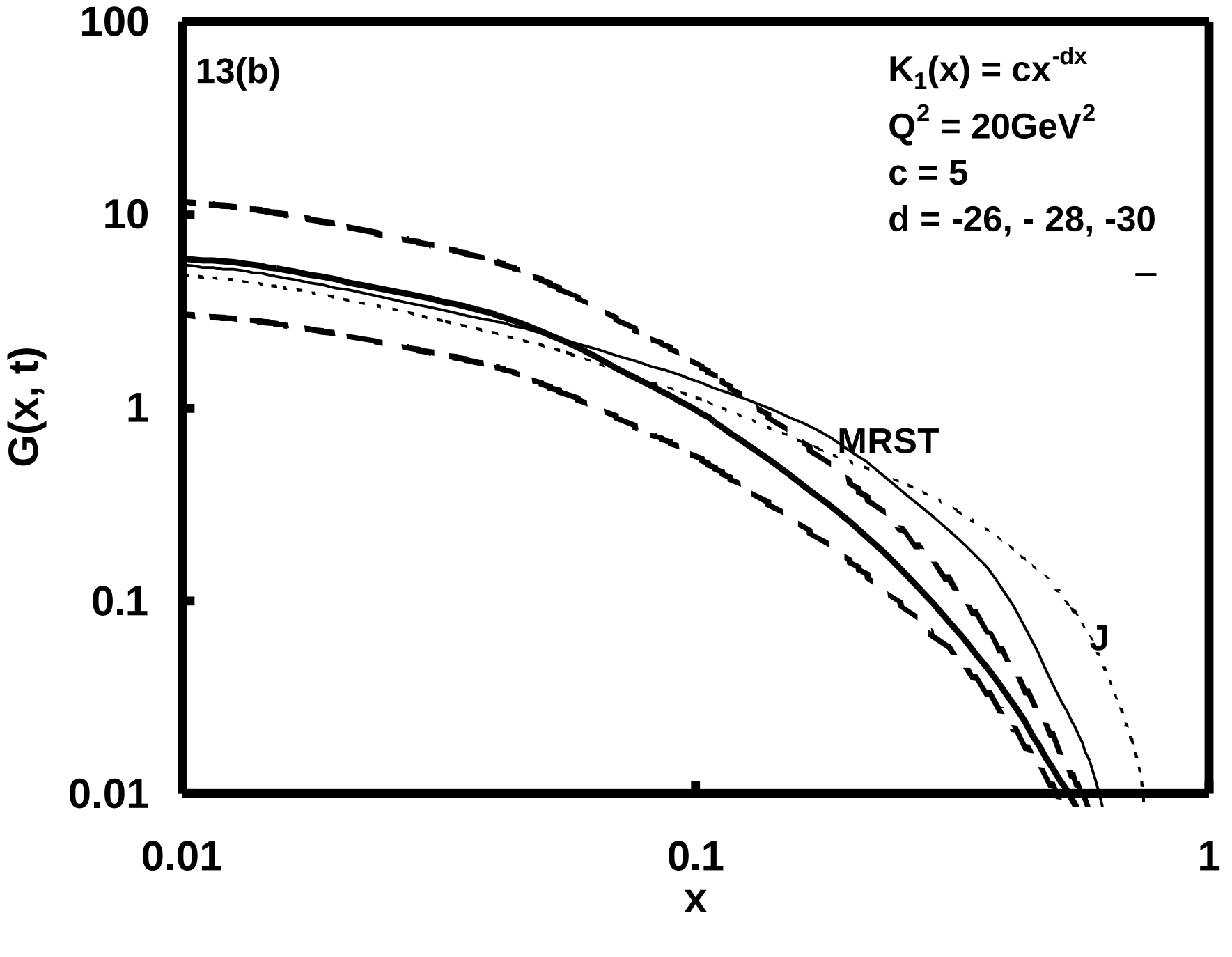

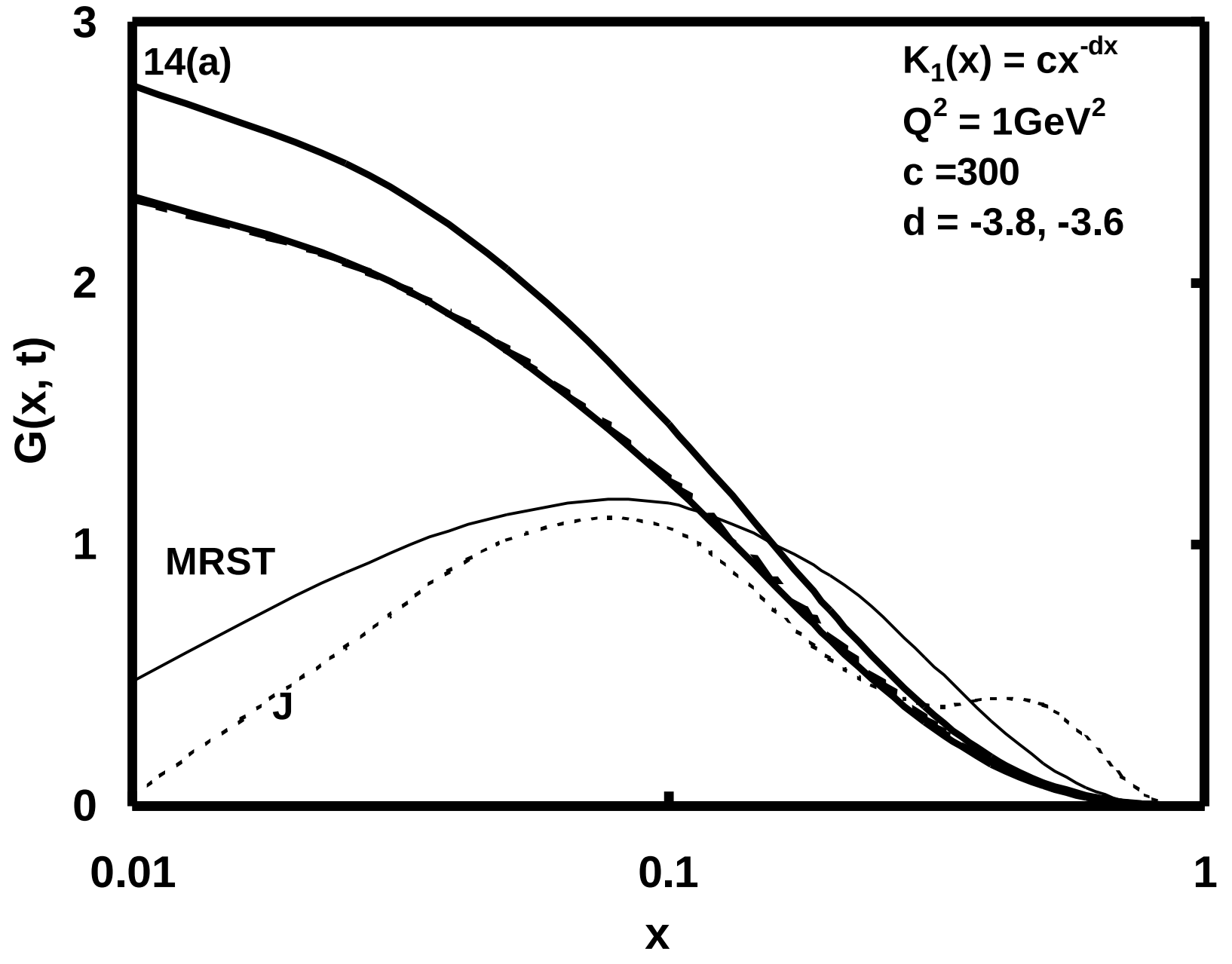

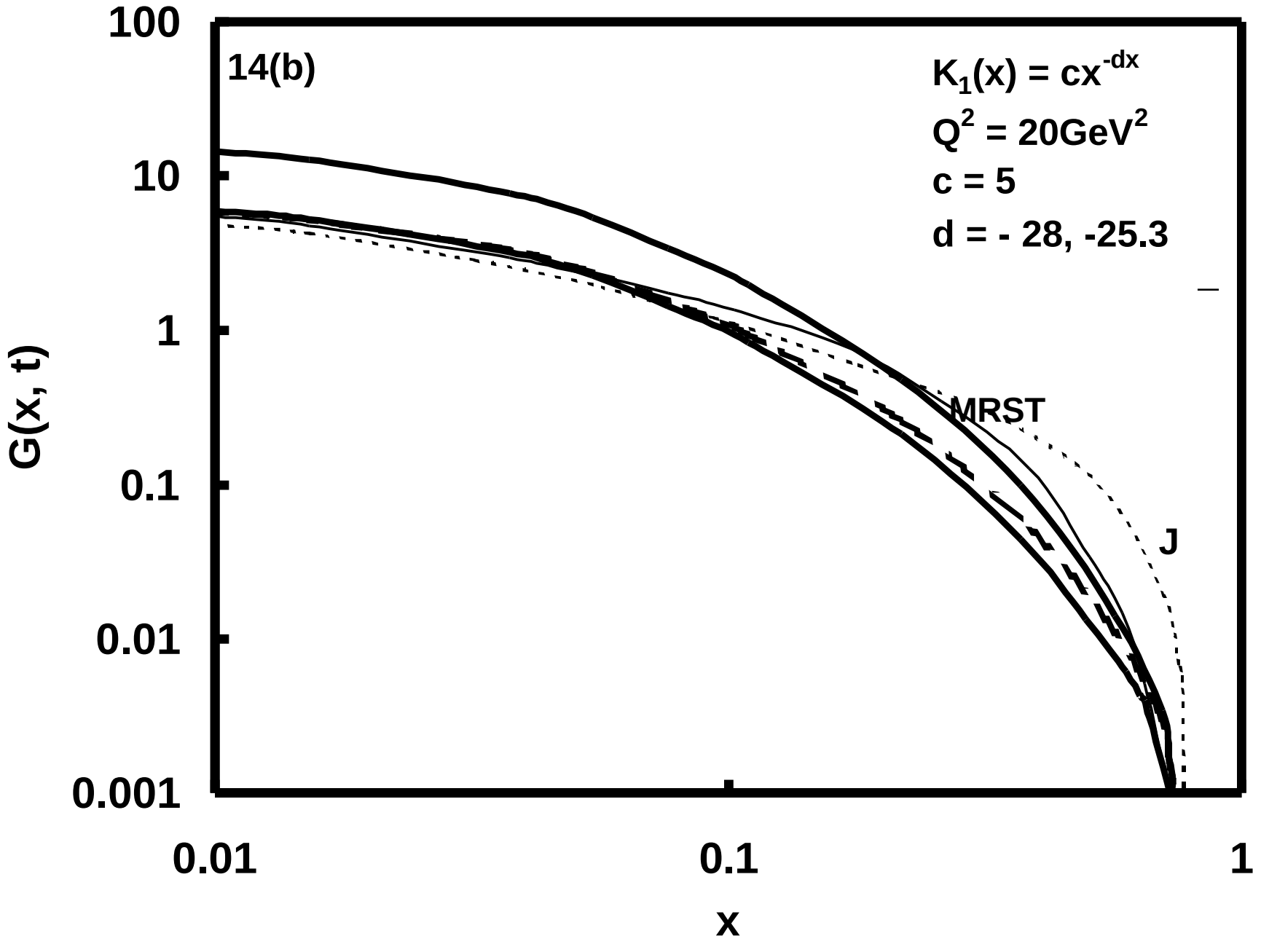

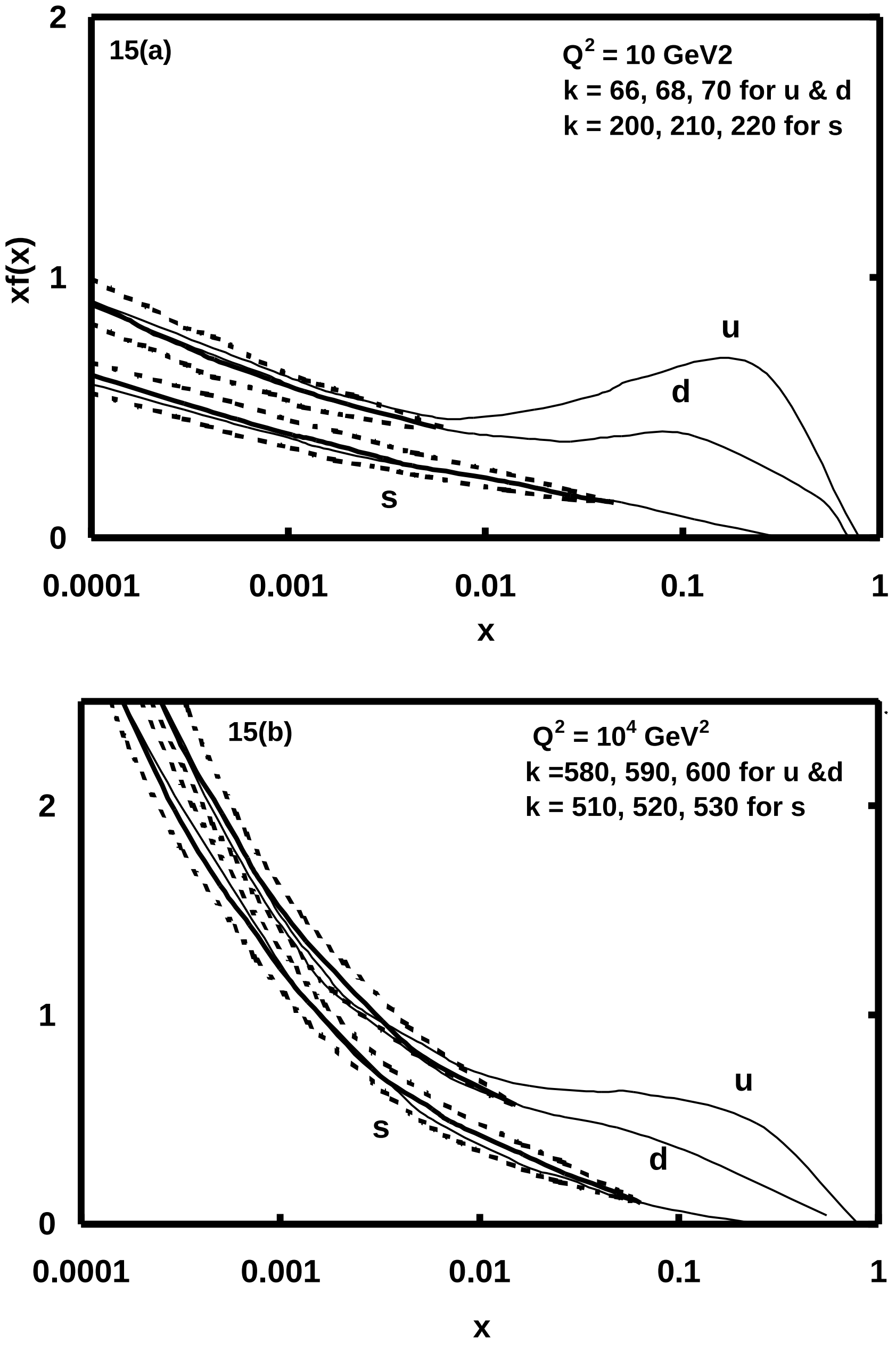

15(a)
Q2 = 10 GeV2
k = 66, 68, 70 for u & d
k = 200, 210, 220 for s
u
d
s
xf(x)
2
1
0
0.0001
0.001
0.01
0.1
1
x
15(b)
Q2 = 104 GeV2
k =580, 590, 600 for u &d
k = 510, 520, 530 for s
u
s
d
xf(x)
2
1
0
0.0001
0.001
0.01
0.1
1
x

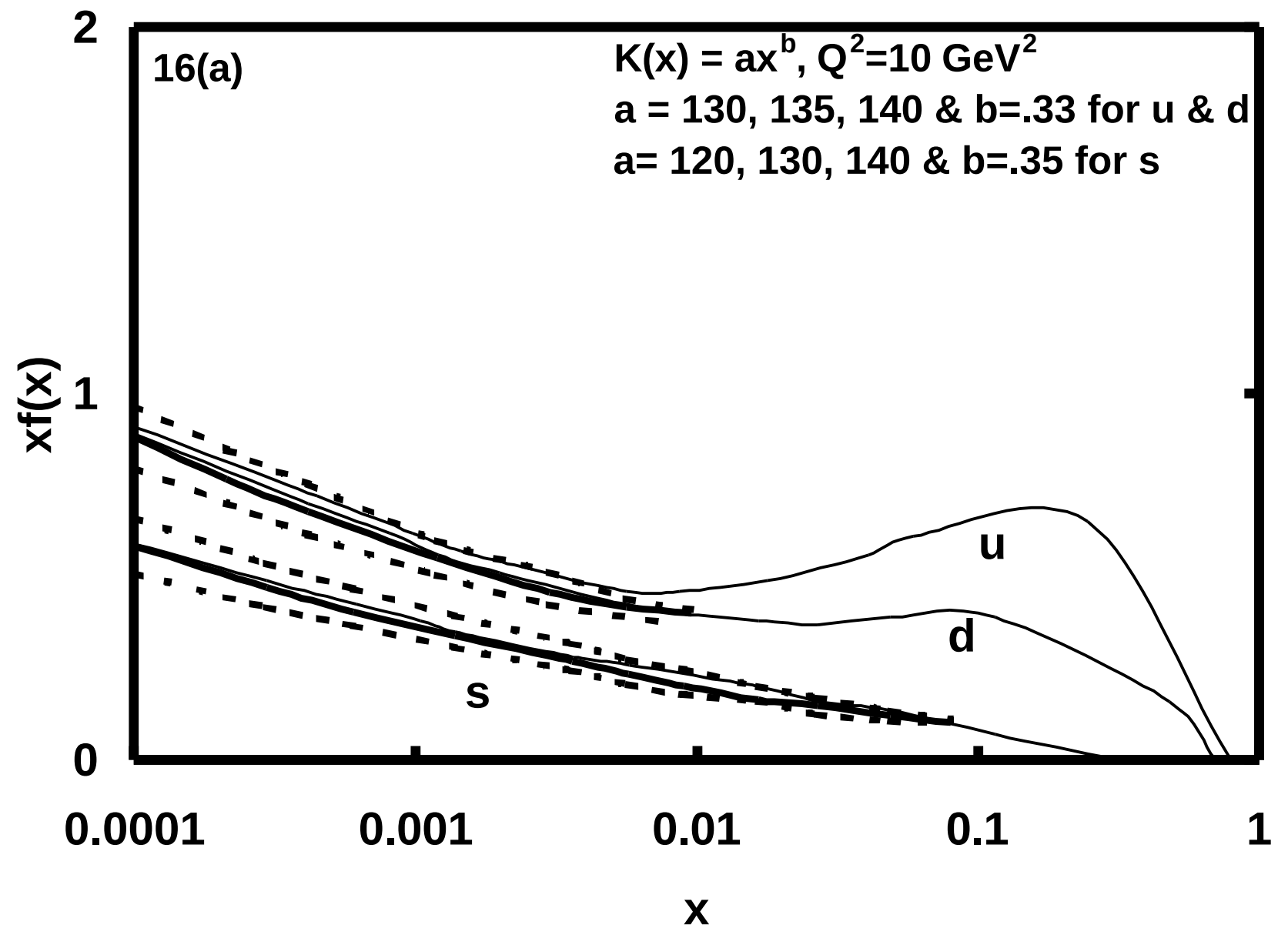
2
16(a)
K(x) = ax^b, Q^2=10 GeV^2
a = 130, 135, 140 & b=.33 for u & d
a= 120, 130, 140 & b=.35 for s
xf(x)
1
u
d
s
0
0.0001
0.001
0.01
0.1
1
x

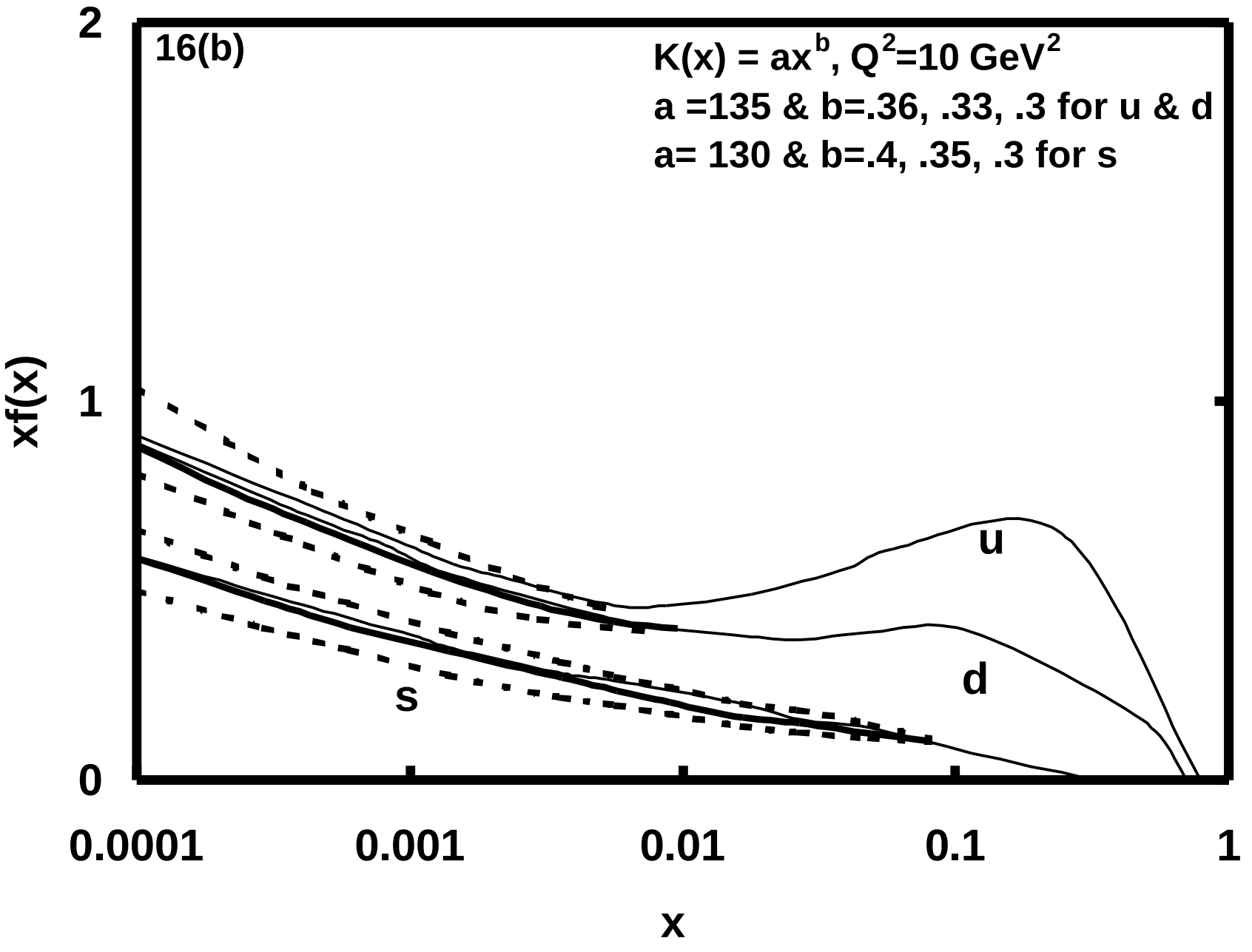
2
16(b)
K(x) = ax^b, Q^2=10 GeV^2
a =135 & b=.36, .33, .3 for u & d
a= 130 & b=.4, .35, .3 for s
xf(x)
1
u
d
s
0
0.0001
0.001
0.01
0.1
1
x

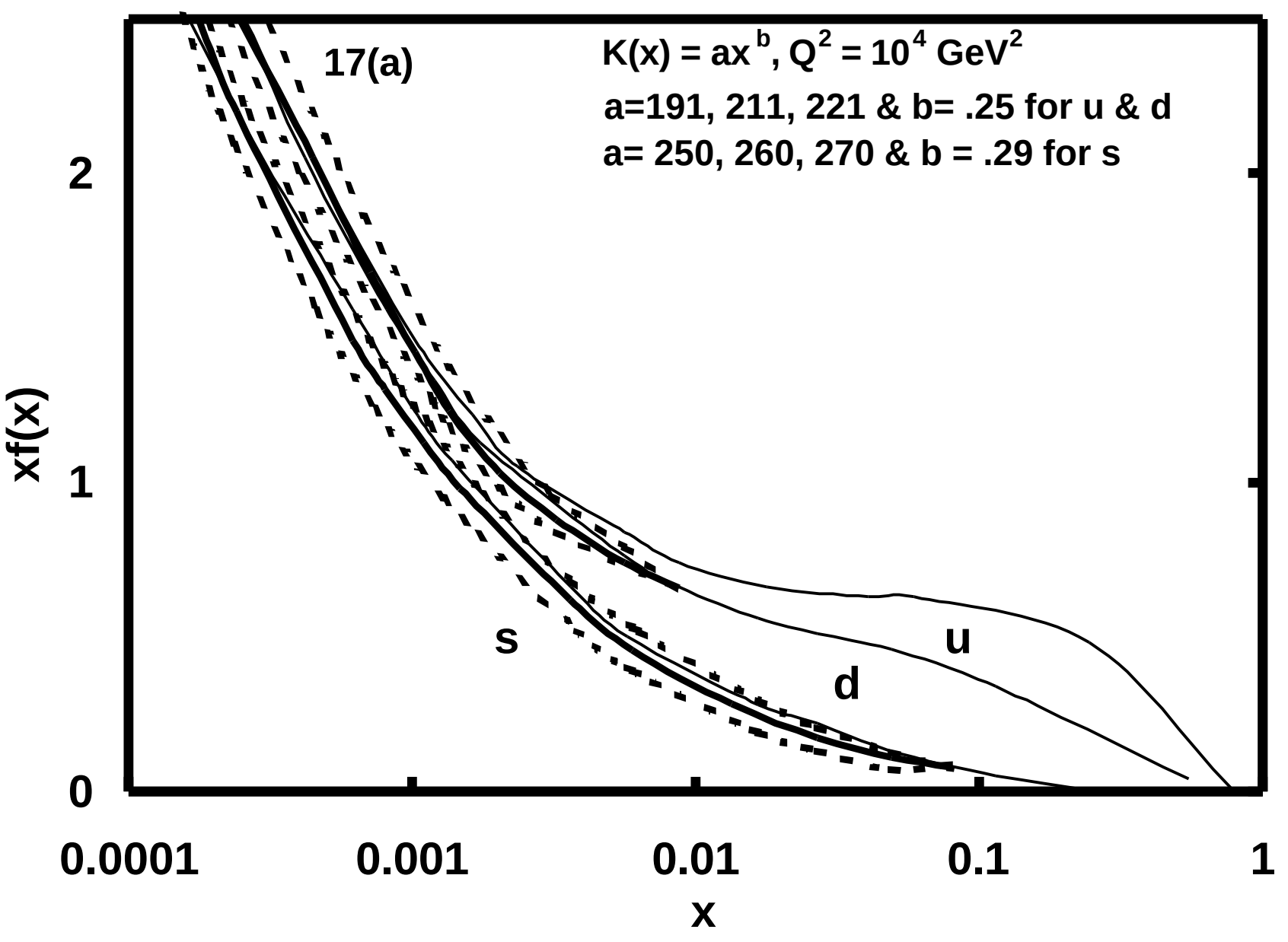
17(a)
K(x) = ax^b, Q^2 = 10^4 GeV^2
a=191, 211, 221 & b= .25 for u & d
a= 250, 260, 270 & b = .29 for s
xf(x)
0
1
2
s
u
d
0.0001
0.001
0.01
0.1
1
x

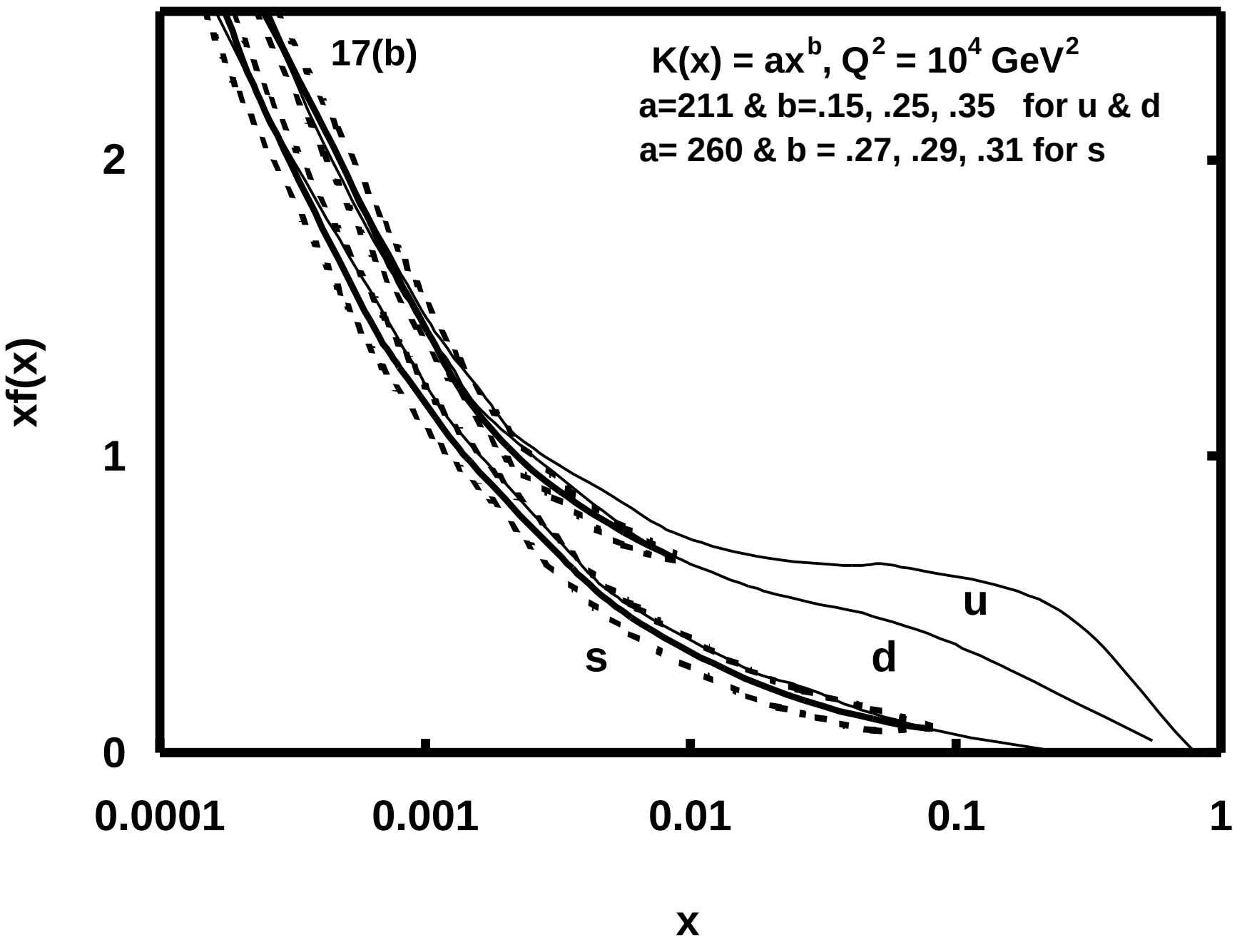
17(b)
K(x) = ax^b, Q^2 = 10^4 GeV^2
a=211 & b=.15, .25, .35 for u & d
a= 260 & b = .27, .29, .31 for s
xf(x)
0
1
2
u
s
d
0.0001
0.001
0.01
0.1
1
x

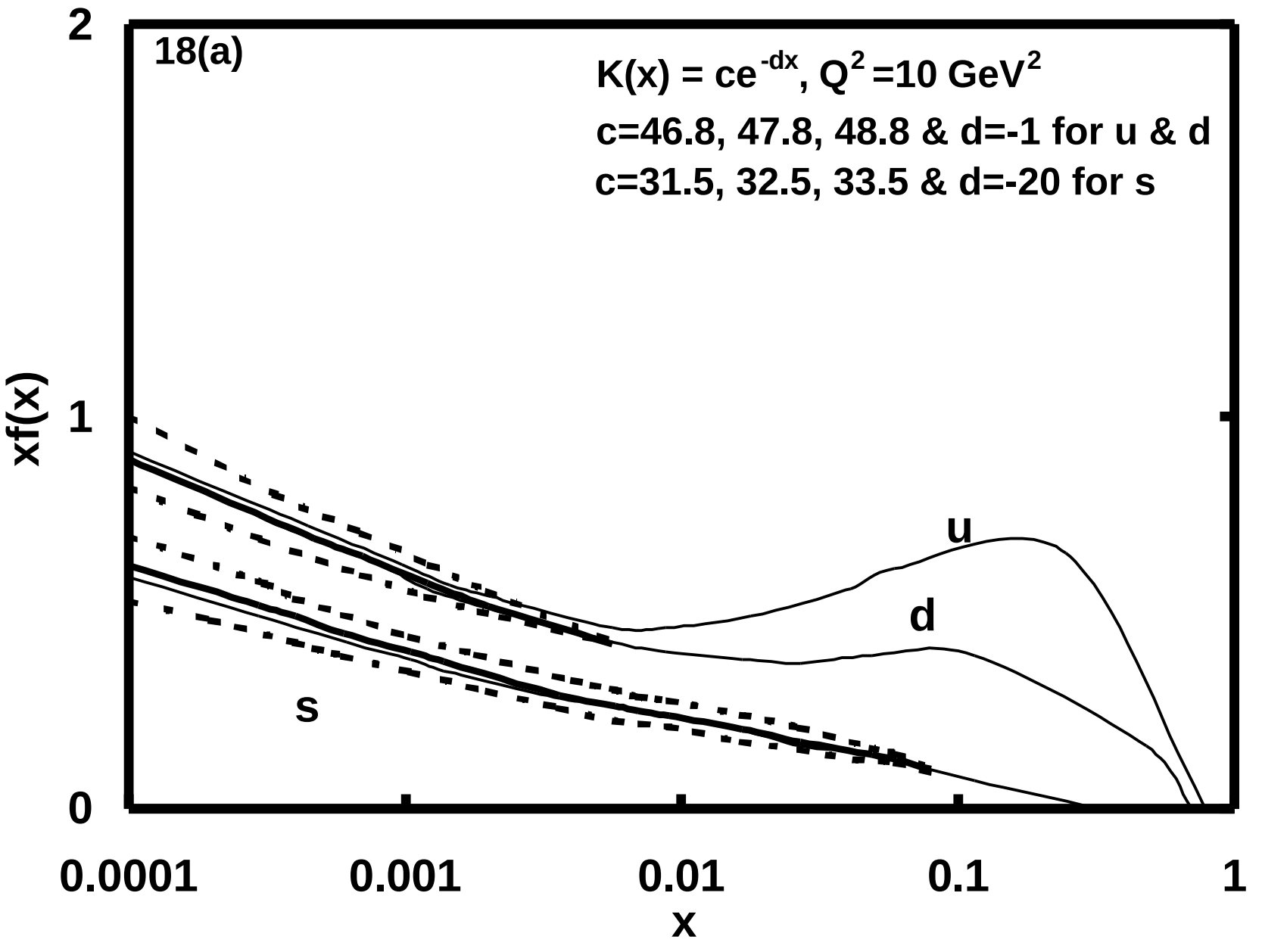

18(a)
K(x) = ce^-dx, Q^2 =10 GeV^2
c=46.8, 47.8, 48.8 & d=-1 for u & d
c=31.5, 32.5, 33.5 & d=-20 for s
xf(x)
u
d
s
x
2
1
0
0.0001
0.001
0.01
0.1
1

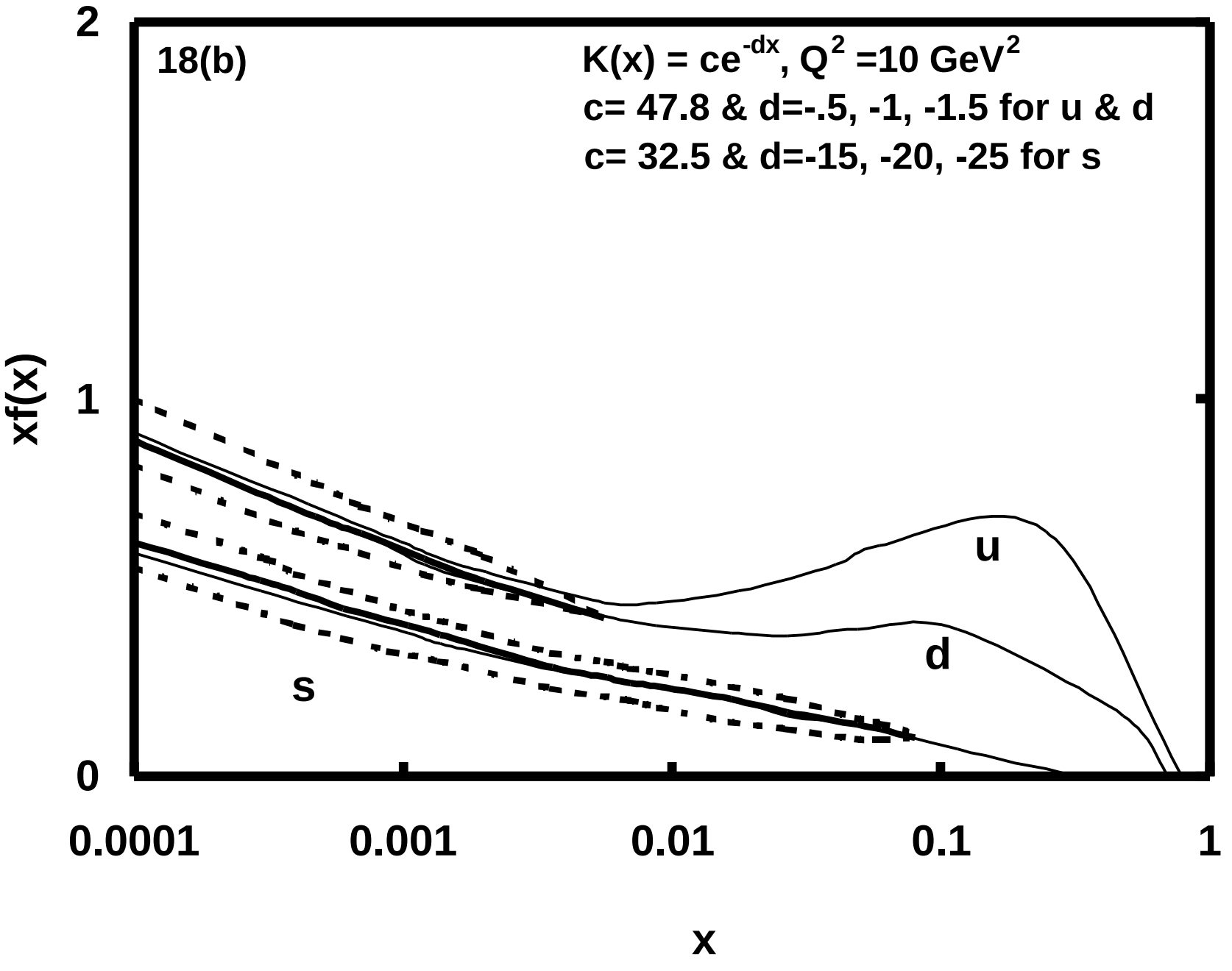

18(b)
K(x) = ce^-dx, Q^2 =10 GeV^2
c= 47.8 & d=-.5, -1, -1.5 for u & d
c= 32.5 & d=-15, -20, -25 for s
xf(x)
u
d
s
x
2
1
0
0.0001
0.001
0.01
0.1
1

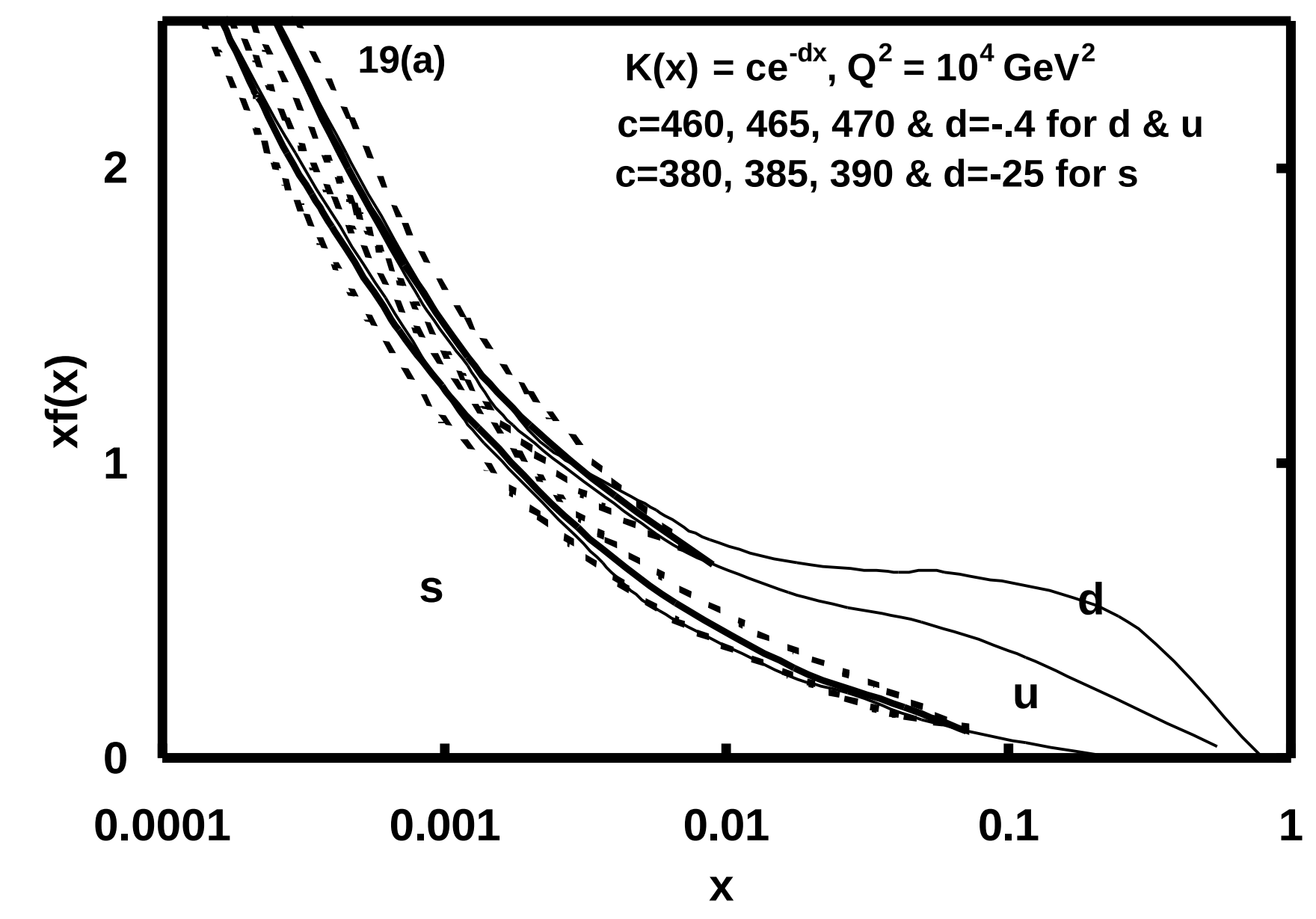
19(a)
K(x) = ce^-dx, Q^2 = 10^4 GeV^2
c=460, 465, 470 & d=-.4 for d & u
c=380, 385, 390 & d=-25 for s
xf(x)
0
1
2
s
d
u
0.0001
0.001
0.01
0.1
1
x

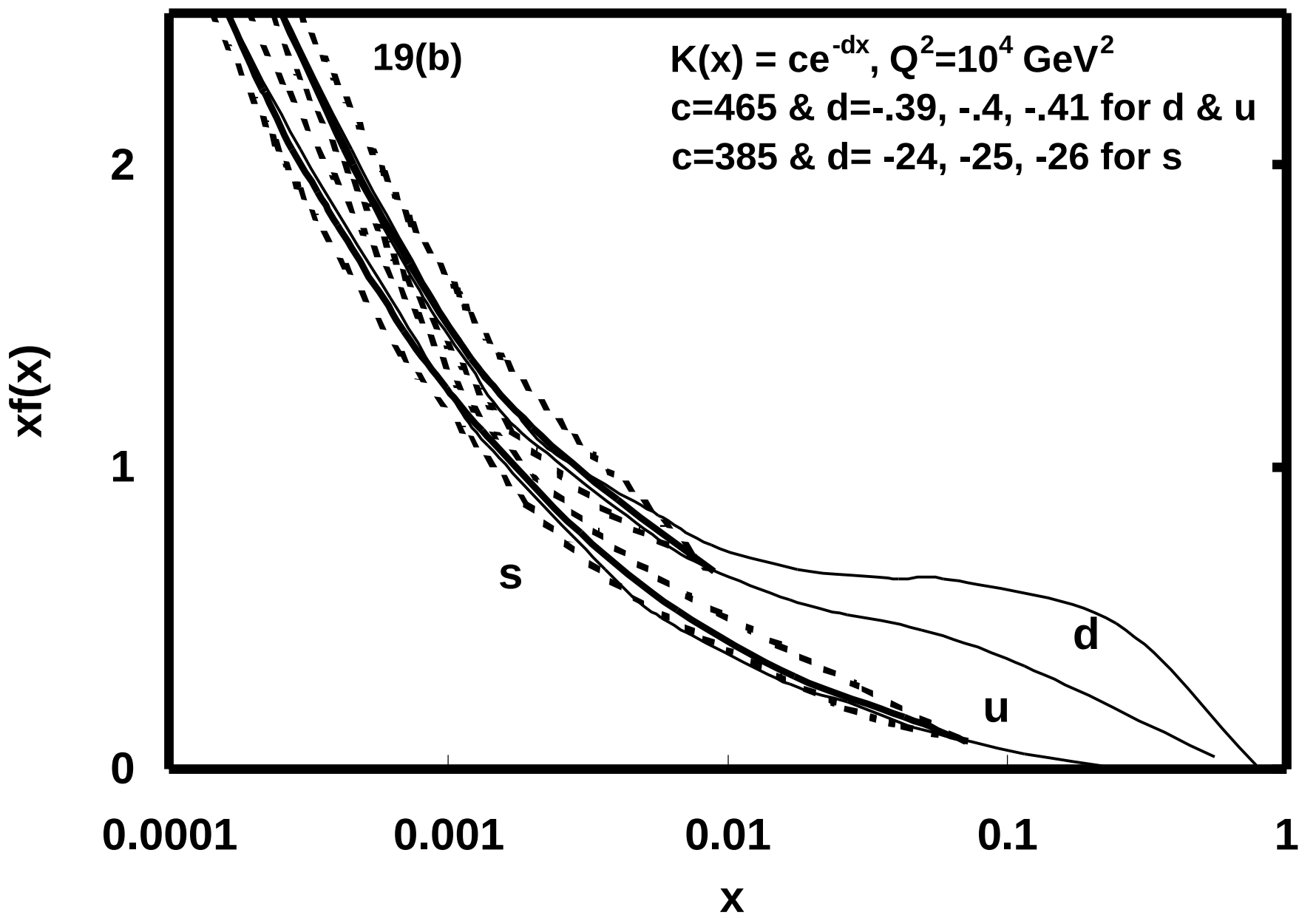
19(b)
K(x) = ce^-dx, Q^2=10^4 GeV^2
c=465 & d=-.39, -.4, -.41 for d & u
c=385 & d= -24, -25, -26 for s
xf(x)
0
1
2
s
d
u
0.0001
0.001
0.01
0.1
1
x